\documentstyle[prc,aps,floats,psfig]{revtex}
\tighten
\begin{document}
\draft

\title{Chiral $2\pi$ exchange at order four and peripheral $NN$ scattering}

\author{D. R. Entem\footnote{On leave from University of Salamanca, Spain.
        E-mail: dentem@uidaho.edu}
        and R. Machleidt\footnote{E-mail: machleid@uidaho.edu}}

\address{Department of Physics, University of Idaho, Moscow, ID 83844, USA}

\date{\today}

\maketitle

\begin{abstract}
We calculate the impact of the complete set of two-pion exchange contributions 
at chiral order four (also known as next-to-next-to-next-to-leading order, N$^3$LO)
on peripheral partial waves of nucleon-nucleon scattering.
Our calculations are based upon the analytical studies by Kaiser.
It turns out that the contribution of order
four is substantially smaller than the one of order three, indicating
convergence of the chiral expansion.
We compare the prediction from chiral pion-exchange with the corresponding
one from conventional meson-theory as represented by the Bonn Full Model
and find, in general, good agreement.
Our calculations provide a sound basis for investigating the issue whether
the low-energy constants determined from $\pi N$ lead to reasonable
predictions for $NN$.
\end{abstract}

\section{Introduction}

One of the most fundamental problems of nuclear physics is to
derive the force between two nucleons from first principles.
A great obstacle for the solution of this problem has been the fact
that the fundamental theory of strong interaction, QCD, is
nonperturbative in the low-energy regime characteristic for
nuclear physics. The way out of this dilemma is paved by the effective
field theory concept which recognizes different energy scales in
nature. Below the chiral symmetry breaking scale,
$\Lambda_\chi \approx 1$ GeV,
the appropriate degrees of freedom are pions and
nucleons interacting via a force that is governed by
the symmetries of QCD, particularly, (broken) chiral symmetry.

The derivation of the nuclear force from chiral effective field
theory was initiated by Weinberg \cite{Wei90} and pioneered
by Ord\'o\~nez \cite{OK92} and van Kolck \cite{ORK94,Kol99}.
Subsequently, many researchers became interested in the field
\cite{CPS92,RR94,KBW97,KGW98,Kai99,Kai01,Kai01a,Kai01b,EGM98,EM01,KSW96,FST97,Par98,Coh97,RS99,Bea01}.
As a result, efficient methods for deriving
the nuclear force from chiral Lagrangians 
emerged~\cite{KBW97,KGW98,Kai99,Kai01,Kai01a,Kai01b}
and the quantitative nature of the chiral nucleon-nucleon ($NN$) potential
improved~\cite{EGM98,EM01}. 

Current $NN$ potentials~\cite{EGM98,EM01} and phase 
shift analyses~\cite{Ren99} include $2\pi$-exchange
contributions up to order three in small momenta
(next-to-next-to-leading order, NNLO). 
However, the contribution at order three is very large,
several times the one at order two (NLO).
This fact raises serious questions concerning the convergence
of the chiral expansion for the two-nucleon problem.
Moreover, it was shown in Ref.~\cite{EM01} that a {\it quantitative}
chiral $NN$ potential requires contact terms of order four.
Consistency then implies that also $2\pi$ (and $3\pi$) contributions
are to be included up to order four.

For the reasons discussed, it is a timely project to investigate
the chiral $2\pi$ exchange contribution to the $NN$ interaction
at order four. Recently, Kaiser~\cite{Kai01a,Kai01b} has derived
the analytic expressions at this order using covariant perturbation
theory and dimensional regularization. It is the chief purpose
of this paper to apply these contributions in peripheral $NN$
scattering and compare the predictions to empirical phase shifts as well as
to the results from conventional meson theory.
Furthermore, we will investigate the above-mentioned convergence issue.
Our calculations provide a sound basis to discuss the question whether
the low-energy constants (LECs) determined from $\pi N$ lead to
reasonable predictions in $NN$.

In Sec.~II, we summarize the Lagrangians involved in the evaluation
of the $2\pi$-exchange contributions presented in Sec.~III.
In Sec.~IV, we explain how we calculate the phase shifts
for peripheral partial waves and present results.
Sec.~V concludes the paper.

\section{Effective chiral Lagrangians}

The effective chiral Lagrangian relevant to our problem
can be written as~\cite{FMS98,Fet00},
\begin{equation}
{\cal L}_{\rm eff} 
=
{\cal L}_{\pi\pi}^{(2)} 
+
{\cal L}_{\pi N}^{(1)} 
+
{\cal L}_{\pi N}^{(2)} 
+
{\cal L}_{\pi N}^{(3)} 
+ \ldots ,
\end{equation}
where the superscript refers to the number of derivatives or 
pion mass insertions (chiral dimension)
and the ellipsis stands for terms of chiral order four or higher.

{\it At lowest/leading order}, 
the $\pi\pi$ Lagrangian is given by,
\begin{equation}
{\cal L}_{\pi\pi}^{(2)} =
\frac{f^2_\pi}{4} \, {\rm tr} \left[ \,
\partial^\mu U \partial_\mu U^\dagger 
+ m_\pi^2 ( U + U^\dagger)\, \right]
\, ,
\label{eq_pipi}
\end{equation}
 and the relativistic $\pi N$ Lagrangian reads,
\begin{equation}
{\cal L}^{(1)}_{\pi N}  =  
 \bar{\Psi} \left(i\gamma^\mu {D}_\mu 
 - M_N
 + \frac{g_A}{2} \gamma^\mu \gamma_5 u_\mu
  \right) \Psi  
\, ,
\end{equation}
with
\begin{eqnarray}
{D}_\mu & = & \partial_\mu + \Gamma_\mu 
\\
\Gamma_\mu & = & \frac12      (
                           \xi^\dagger \partial_\mu \xi
                        +  \xi \partial_\mu \xi^\dagger)
= 
\frac{i}{4f^2_\pi} \,
\mbox{\boldmath $\tau$} \cdot 
 ( \mbox{\boldmath $\pi$}
\times
 \partial_\mu \mbox{\boldmath $\pi$})
+ \ldots
\\
u_\mu & = & i (
                           \xi^\dagger \partial_\mu \xi
                        -  \xi \partial_\mu \xi^\dagger)  
= -
\frac{1}{f_\pi} \,
\mbox{\boldmath $\tau$} \cdot 
 \partial_\mu \mbox{\boldmath $\pi$}
+ \ldots
\\
U & = & \xi^2 =
 1 + 
\frac{i}{f_\pi}
\mbox{\boldmath $\tau$} \cdot \mbox{\boldmath $\pi$}
-\frac{1}{2f_\pi^2} 
\mbox{\boldmath $\pi$}^2
-\frac{i\alpha}{f_\pi^3}
(\mbox{\boldmath $\tau$} \cdot \mbox{\boldmath $\pi$})^3
+\frac{8\alpha-1}{8f_\pi^4} 
\mbox{\boldmath $\pi$}^4
+ \ldots
\label{eq_alpha}
\end{eqnarray}
The coefficient $\alpha$ that appears in the last equation
is arbitrary. Therefore, diagrams with chiral vertices that
involve three or four pions must always be grouped together
such that the $\alpha$-dependence drops out
(cf.\ Fig.~\ref{fig_diag3}, below).

In the above equations, $M_N$ denotes the nucleon mass,
$g_A$ the axial-vector coupling constant,
and $f_\pi$ the pion decay constant.
Numerical values are given in Table~I.

We apply
the heavy baryon (HB) formulation of chiral perturbation theory\cite{BKM95}
in which the relativistic $\pi N$ Lagrangian is subjected
to an expansion in terms of powers of $1/M_N$ (kind of a
nonrelativistic expansion), the lowest order of which is 
\begin{eqnarray}
\widehat{\cal L}^{(1)}_{\pi N} & 
= & 
\bar{N} \left(
 i {D}_0 
 - \frac{g_A}{2} \; 
\vec \sigma \cdot \vec u
\right) N  
\nonumber \\
 & 
= & 
\bar{N} \left[ i \partial_0 
- \frac{1}{4f_\pi^2} \;
\mbox{\boldmath $\tau$} \cdot 
 ( \mbox{\boldmath $\pi$}
\times
 \partial_0 \mbox{\boldmath $\pi$})
- \frac{g_A}{2f_\pi} \;
\mbox{\boldmath $\tau$} \cdot 
 ( \vec \sigma \cdot \vec \nabla )
\mbox{\boldmath $\pi$} \right] N + \ldots
\label{eq_L1}
\end{eqnarray}
In the relativistic formulation, the field operators
representing nucleons, $\Psi$, contain 
four-component Dirac spinors; 
while in the HB version, the field operators, $N$,
contain Pauli spinors; in addition,
all nucleon field operators contain
Pauli spinors describing the isospin of the nucleon.

{\it At dimension two},
the relativistic $\pi N$ Lagrangian reads
\begin{equation}
{\cal L}^{(2)}_{\pi N} 
= \sum_{i=1}^{4} c_i \bar{\Psi} O^{(2)}_i \Psi \, .
\label{eq_L2rel}
\end{equation}
The various operators $O^{(2)}_i$ are given in Ref.\cite{Fet00}.
The fundamental rule by which this Lagrangian---as well as all the other
ones---are assembled is that they must contain {\it all\/} terms
consistent with chiral symmetry and Lorentz invariance (apart from the other trivial
symmetries) at a given chiral dimension (here: order two).
The parameters $c_i$ are known as low-enery constants (LECs)
and are determined empirically from fits to $\pi N$ data (Table~I).

The HB projected $\pi N$ Lagrangian at order two is most conveniently broken up
into two pieces,
\begin{equation}
\widehat{\cal L}^{(2)}_{\pi N} \, = \,
\widehat{\cal L}^{(2)}_{\pi N, \, \rm fix} \, + \,
\widehat{\cal L}^{(2)}_{\pi N, \, \rm ct} \, ,
\label{eq_L2}
\end{equation}
with
\begin{equation}
\widehat{\cal L}^{(2)}_{\pi N, \, \rm fix}  =  
 \bar{N} \left[
\frac{1}{2M_N}\: \vec D \cdot \vec D
+ i\, \frac{g_A}{4M_N}\: \{\vec \sigma \cdot \vec D, u_0\}
 \right] N
\label{eq_L2fix}
\end{equation}
and
\begin{eqnarray}
\widehat{\cal L}^{(2)}_{\pi N, \, \rm ct}
& = & 
 \bar{N} \left[
 2\,
c_1
\, m_\pi^2\, (U+U^\dagger)
\, + \, \left( 
c_2
- \frac{g_A^2}{8M_N}\right) u_0^2
 \, + \,
c_3
\, u_\mu  u^\mu
%\right.  \nonumber \\ && \left.
+ \, \frac{i}{2} \left( 
c_4
+ \frac{1}{4M_N} \right) 
  \vec \sigma \cdot ( \vec u \times \vec u)
 \right] N \, .
\label{eq_L2ct}
\end{eqnarray}
Note that 
$\widehat{\cal L}^{(2)}_{\pi N, \, \rm fix}$  
is created entirely from the HB expansion of the relativistic
${\cal L}^{(1)}_{\pi N}$ and thus has no free parameters (``fixed''),
while 
$\widehat{\cal L}^{(2)}_{\pi N, \, \rm ct}$
is dominated by the new $\pi N$ contact terms proportional to the
$c_i$ parameters, besides some small $1/M_N$ corrections.

{\it At dimension three},
the relativistic $\pi N$ Lagrangian can be formally written as
\begin{equation}
{\cal L}^{(3)}_{\pi N} = \sum_{i=1}^{23} d_i \bar{\Psi} O^{(3)}_i \Psi \, ,
\label{eq_L3rel}
\end{equation}
with the operators, $O^{(3)}_i$, listed in Refs.\cite{FMS98,Fet00}; 
not all 23 terms are of interest here.
The new LECs that occur at this order are the $d_i$.
Similar to the order two case,
the HB projected Lagrangian at order three can be broken into two pieces,
\begin{equation}
\widehat{\cal L}^{(3)}_{\pi N} \, = \,
\widehat{\cal L}^{(3)}_{\pi N, \, \rm fix} \, + \,
\widehat{\cal L}^{(3)}_{\pi N, \, \rm ct} \, ,
\label{eq_L3}
\end{equation}
with
$\widehat{\cal L}^{(3)}_{\pi N, \, \rm fix}$
and
$\widehat{\cal L}^{(3)}_{\pi N, \, \rm ct}$
given in Refs.\cite{FMS98,Fet00}.

\section{Noniterative $2\pi$ exchange contributions to the $NN$ interaction}

The effective Lagrangian presented in the previous section
is the crucial ingredient for the evaluation of
the pion-exchange contributions to the nucleon-nucleon ($NN$) interaction.
Since we are dealing here with a low-energy
effective theory, it is appropriate to analyze the diagrams
in terms of powers of small momenta: $(Q/\Lambda_\chi)^\nu$,
where $Q$ stands for a momentum (nucleon three-momentum or
pion four-momentum) or a pion mass and $\Lambda_\chi \approx 1$ GeV
is the chiral symmetry breaking scale.
This procedure has become known as power counting.
For non-iterative contributions to the $NN$ interaction
(i.~e., irreducible graphs with four external nucleon legs), 
the power $\nu$ of a diagram is given by
\begin{equation}
\nu = 2 \, l \, + \sum_j \left( d_j + \frac{n_j}{2} - 2 \right) \, ,
\end{equation}
where $l$ denotes the number of loops in the diagram,
$d_j$ the number of derivatives or pion-mass insertions 
and
$n_i$ the number of nucleon fields 
involved in vertex $j$;
the sum runs over all vertices $j$ contained in the diagram under consideration.

At order zero ($\nu=0$, lowest order, leading order, LO), 
we have only the static one-pion-exchange (OPE)
and, at order one, there are no pion-exchange contributions.
Higher order graphs are shown in Figs.~\ref{fig_diag1}-\ref{fig_diag3}.
Analytic results for these graphs were derived by Kaiser
and coworkers~\cite{KBW97,Kai01a,Kai01b} 
using covariant perturbation, i.~e., they start
out with the relativistic versions of the $\pi N$ Lagrangians
(see previous section).
Relativistic vertices and nucleon propagators are then 
expanded in powers of $1/M_N$. 
The divergences that occur in conjunction with the four-dimensional
loop integrals are treated by means of dimensional regularization,
a prescription which is
consistent with chiral symmetry and power counting.
The results derived in this way
are the same obtained when starting right away
with the HB versions of the $\pi N$ Lagrangians.
However, as it turns out, the method used by
the Munich group is more efficient in dealing 
with the rather tedious calculations.

We will state the analytical results in terms of contributions to the 
on-shell momentum-space $NN$ amplitude
which has the general form,
\begin{eqnarray} 
V({\vec p}~', \vec p)  = 
\frac{1}{(2\pi)^3} 
& \bigg\{ &
\:\, V_C \,\,
+ \bbox{\tau}_1 \cdot \bbox{\tau}_2 \, 
W_C 
\nonumber \\ &+&  
\left[ \, V_S \:\, + \bbox{\tau}_1 \cdot \bbox{\tau}_2 \, W_S \,\:\, \right] \,
\vec\sigma_1 \cdot \vec \sigma_2
\nonumber \\ &+& 
\left[ \, V_T \:\,     + \bbox{\tau}_1 \cdot \bbox{\tau}_2 \, W_T \,\:\, \right] \,
\vec \sigma_1 \cdot \vec q \,\, \vec \sigma_2 \cdot \vec q  
\nonumber \\ &+& 
\left[ \, V_{LS} + \bbox{\tau}_1 \cdot \bbox{\tau}_2 \, W_{LS}    \right] \,
\left(-i \vec S \cdot (\vec q \times \vec k) \,\right)
\nonumber \\ &+& 
\left[ \, V_{\sigma L} + \bbox{\tau}_1 \cdot \bbox{\tau}_2 \, W_{\sigma L} \, \right] \,
\vec\sigma_1\cdot(\vec q\times \vec k\,) \,\,
\vec \sigma_2 \cdot(\vec q\times \vec k\,)
\bigg\}
\, ,
\label{eq_nnamp}
\end{eqnarray}
where ${\vec p}~'$ and $\vec p$ denote the final and initial nucleon momentum 
in the center-of-mass (CM) frame, respectively,
\begin{displaymath}
\begin{array}{llll}
\vec q &\equiv& {\vec p}~' - \vec p &  \mbox{\rm is the momentum transfer},\\
\vec k &\equiv& \frac12 ({\vec p}~' + \vec p) & \mbox{\rm the average momentum},\\
\vec S &\equiv& \frac12 (\vec\sigma_1+\vec\sigma_2) & \mbox{\rm the total spin},
\end{array}
\end{displaymath}
and $\vec \sigma_{1,2}$ and $\bbox{\tau}_{1,2}$ are the spin and isospin 
operators, respectively, of nucleon 1 and 2.
For on-energy-shell scattering, $V_\alpha$ and $W_\alpha$ 
($\alpha=C,S,T,LS,\sigma L$) can be expressed as functions of $q$ and $k$ (with
$q\equiv |\vec q|$ and $k\equiv |\vec k|$), only.

Our formalism is similar to the one used in 
Refs.~\cite{KBW97,KGW98,Kai99,Kai01,Kai01a,Kai01b}, except for two
differences: all our momentum space amplitudes
differ by an over-all factor of $(-1)$ and our spin-orbit amplitudes,
$V_{LS}$ and $W_{LS}$, differ by an additional factor of $(-2)$ from
the conventions used by Kaiser 
{\it et al.}~\cite{KBW97,KGW98,Kai99,Kai01,Kai01a,Kai01b}.
We have chosen our conventions such that they are closely
in tune with what is commonly used in nuclear physics.

We stress that, throughout this paper, we consider {\it on-shell} $NN$
amplitudes, i.~e., we always assume $|{\vec p}~'|=|\vec p| \equiv p$.
Note also that we will state only the {\it nonpolynomial} part of the
amplitudes.
Polynomial terms can be absorbed into contact interactions that are not the
subject of this study. Moreover, in Sec.~IV, below,
we will show results for $NN$ scattring in $F$ and higher partial waves
(orbital angular momentum $L\geq 3$) where polynomials of order $Q^\nu$ with
$\nu\leq 4$ do not contribute.

\subsection{Order two}

Two-pion exchange occurs first at order two ($\nu=2$,
next-to-leading order, NLO), also know as leading-order $2\pi$ exchange. 
The graphs are shown in the first row of 
Fig.~\ref{fig_diag1}.
Since a loop creates already $\nu=2$,
the vertices involved at this order can only be from the leading/lowest order
Lagrangian $\widehat{\cal L}^{(1)}_{\pi N}$, Eq.~(\ref{eq_L1}),
i.~e., they carry only one derivative.
These vertices are denoted by small dots in
Fig.~\ref{fig_diag1}.
Note that, here, we include only the non-iterative part
of the box diagram which is obtained by subtracting the
iterated OPE contribution 
[Eq.~(\ref{eq_2piit}), below, but using
$M_N^2/E_p \approx M_N$]
from the full box diagram at order two.
To make this paper selfcontained and to uniquely define
the contributions for which we will show results in Sec.~IV, below,
we summarize here the explicit mathematical expressions
derived in Ref.~\cite{KBW97}:
\begin{eqnarray} 
W_C &=&-{L(q)\over384\pi^2 f_\pi^4} 
\left[4m_\pi^2(5g_A^4-4g_A^2-1)
+q^2(23g_A^4 -10g_A^2-1) 
+ {48g_A^4 m_\pi^4 \over w^2} \right] \,,  
\label{eq_2C}
\\   
V_T &=& -{1\over q^2} V_{S} = -{3g_A^4 L(q)\over 64\pi^2 f_\pi^4} \,, 
\label{eq_2T}
\end{eqnarray}  
where
\begin{equation} 
L(q) \equiv{w\over q} \ln {w+q \over 2m_\pi}
\end{equation}
and
\begin{equation} 
 w\equiv \sqrt{4m_\pi^2+q^2} \,. 
\end{equation}

\subsection{Order three}

The two-pion exchange diagrams of order three ($\nu=3$, next-to-next-to-leading order,
NNLO) are very similar to the ones of order two, except that 
they contain one insertion from 
$\widehat{\cal L}^{(2)}_{\pi N}$, Eq.~(\ref{eq_L2}). 
The resulting contributions are typically either proportional to one of the 
low-energy constants $c_i$ or they contain a factor $1/M_N$.
Notice that relativistic $1/M_N$ corrections can occur
for vertices and nucleon propagators.
In Fig.~\ref{fig_diag1}, we show in row two the diagrams with
vertices proportional to $c_i$ (large solid dot), Eq.~(\ref{eq_L2ct}),
and in row three and four a few representative graphs with a $1/M_N$ 
correction (symbols with an open circle). The number of $1/M_N$ correction graphs
is large and not all are shown in the figure.
Again, the box diagram is corrected for a contribution from
the iterated OPE: in Eq.~(\ref{eq_2piit}), below, the expansion of
the factor $M^2_N/E_p = M_N - p^2/M_N + \ldots$ is applied;
the term proportional to $(-p^2/M_N)$ is subtracted from
the third order box diagram contribution.
For completeness, we recall here the mathematical expressions
derived in Ref.~\cite{KBW97}.
\begin{eqnarray} 
V_C &=&{3g_A^2 \over 16\pi f_\pi^4} 
\left\{ 
 {g_A^2 m_\pi^5  \over 16M_N w^2}  
-\left[2m_\pi^2( 2c_1-c_3)-q^2  \left(c_3 +{3g_A^2\over16M_N}\right)\right]
\widetilde{w}^2 A(q) \right\} \,, 
\label{eq_3C}
\\
W_C &=& {g_A^2\over128\pi M_N f_\pi^4} \left\{ 
 3g_A^2 m_\pi^5 w^{-2} 
 - \left[ 4m_\pi^2 +2q^2-g_A^2(4m_\pi^2+3q^2) \right] 
\widetilde{w}^2 A(q)
\right\} 
\,,\\ 
V_T &=& -{1 \over q^2} V_{S}=
{9g_A^4 \widetilde{w}^2 A(q) \over 512\pi M_N f_\pi^4} 
 \,,  \\ 
W_T &=&-{1\over q^2}W_{S} =-{g_A^2 A(q) \over 32\pi f_\pi^4}
\left[
\left( c_4 +{1\over 4M_N} \right) w^2
-{g_A^2 \over 8M_N} (10m_\pi^2+3q^2)  \right] 
\,,\\
V_{LS} &=&  {3g_A^4  \widetilde{w}^2 A(q) \over 32\pi M_N f_\pi^4} 
 \,,\\  
W_{LS} &=& {g_A^2(1-g_A^2)\over 32\pi M_N f_\pi^4} 
w^2 A(q) \,, 
\label{eq_3LS}
\end{eqnarray}   
with
\begin{equation} 
A(q) \equiv {1\over 2q}\arctan{q \over 2m_\pi} 
\end{equation}
and
\begin{equation} 
\widetilde{w} \equiv  \sqrt{2m_\pi^2+q^2} \,. 
\end{equation}

\subsection{Order four}

This order, which may also be denoted by
next-to-next-to-next-to-leading order (N$^3$LO), is the main focus of this paper.
There are one-loop graphs 
(Fig.~\ref{fig_diag2}) 
and
two-loop contributions 
(Fig.~\ref{fig_diag3}).

\subsubsection{One-loop diagrams}

\paragraph{$c_i^2$ contributions}
The only contribution of this kind comes from the football diagram with both
vertices proportional to $c_i$ (first row of
Fig.~\ref{fig_diag2}). One obtains~\cite{Kai01a}:
\begin{eqnarray}  
V_C & = & {3 L(q) \over 16 \pi^2 f_\pi^4 } 
\left[
\left( {c_2 \over 6} w^2 +c_3 \widetilde{w}^2 -4c_1 m_\pi^2 \right)^2 
+{c_2^2 \over 45 } w^4 
\right] \,, 
\label{eq_4c2C}
\\
W_T  &=&  -{1\over q^2} W_S 
= {c_4^2 w^2 L(q) \over 96 \pi^2 f_\pi^4 } 
\,. 
\label{eq_4c2T}
\end{eqnarray}

\paragraph{$c_i/M_N$ contributions}
This class consists of diagrams with one vertex proportional to $c_i$
and one $1/M_N$ correction.
A few graphs that are representative for this class are shown in the
second row of Fig.~\ref{fig_diag2}. 
Symbols with a large solid dot and an open circle denote 
$1/M_N$ corrections of vertices
proportional to $c_i$. They are part of
$\widehat{\cal L}^{(3)}_{\pi N}$, Eq.~(\ref{eq_L3}).
The result for this group of diagrams is~\cite{Kai01a}:
\begin{eqnarray} 
V_C & = & - {g_A^2\, L(q) \over 32 \pi^2 M_N f_\pi^4 } \left[ 
(c_2-6c_3) q^4 +4(6c_1+c_2-3c_3)q^2 m_\pi^2 
+6(c_2-2c_3)m_\pi^4
%\right.  \nonumber \\ && \left.
+24(2c_1+c_3)m_\pi^6 w^{-2} \right] \,,
\label{eq_4cMC}
\\
W_C &=& 
-{c_4 q^2 L(q) \over 192 \pi^2 M_N f_\pi^4 } 
\left[ g_A^2 (8m_\pi^2+5q^2) + w^2 \right] 
\,, \\
W_T  &=&  -{1\over q^2} W_S = 
-{c_4 L(q) \over 192 \pi^2 M_N f_\pi^4 } 
\left[ g_A^2 (16m_\pi^2+7q^2) - w^2 \right] 
\label{eq_4cMS}
\,,  \\
V_{LS}& = & {c_2 \, g_A^2 \over 8 \pi^2 M_N f_\pi^4 } 
\, w^2 L(q) 
\,, \\
W_{LS}  &=& 
-{c_4 L(q) \over 48 \pi^2 M_N f_\pi^4 } 
\left[ g_A^2 (8m_\pi^2+5q^2) + w^2 \right] 
\,.
\label{eq_4cMLS}
\end{eqnarray}

\paragraph{$1/M_N^2$ corrections}
These are relativistic $1/M_N^2$ corrections of the leading order
$2\pi$ exchange diagrams. Typical examples
for this large class are shown in row three to six of
Fig.~\ref{fig_diag2}. 
This time, there is no correction from the iterated OPE, Eq.~(\ref{eq_2piit}),
since the expansion of the factor $M^2_N/E_p$ does not create
a term proportional to $1/M^2_N$. 
The total result for this class is~\cite{Kai01b},
\begin{eqnarray} 
V_C &=& -{g_A^4 \over 32\pi^2 M_N^2 f_\pi^4}
\Bigg[ 
L(q) \, \Big(2m_\pi^8 w^{-4}+8m_\pi^6 w^{-2} -q^4 -2m_\pi^4\Big)
+{ m_\pi^6 \over 2 w^{2}}\, 
\Bigg] \,,
\label{eq_4M2C}
\\
W_C           &=& -{1\over 768\pi^2 M_N^2 f_\pi^4} \Bigg\{ L(q) \, \bigg[
%\nonumber \\  && 
                8g_A^2 \, \bigg({3\over 2} q^4 +3m_\pi^2 q^2 +3m_\pi^4
               -6m_\pi^6 w^{-2} -k^2(8m_\pi^2 +5q^2) \bigg)
\nonumber \\   && 
                + 4g_A^4 
                \bigg(k^2\big(20m_\pi^2+7q^2-16m_\pi^4 w^{-2}\big) 
                +16m_\pi^8 w^{-4}+12 m_\pi^6 w^{-2} 
%\nonumber \\   && 
-4m_\pi^4q^2w^{-2} -5q^4 -6m_\pi^2 q^2-6m_\pi^4 \bigg) 
%\nonumber \\   && 
               -4k^2 w^2 \, \bigg]
\nonumber \\   && 
                \,+\, {16 g_A^4 m_\pi^6 \over w^{2}} \Bigg\} \,,
\\
V_T &=& -{1\over q^2} V_S= {g_A^4 \, L(q) \over 32\pi^2 M_N^2 f_\pi^4} 
\bigg(k^2+{5\over 8} q^2 +m_\pi^4 w^{-2} \bigg) \,,
\\
W_T &=& -{1\over q^2} W_S = { L(q) \over 1536\pi^2 M_N^2 f_\pi^4} 
\Bigg[\, 4 g_A^4\, \bigg( 7m_\pi^2+{17\over 4} q^2 +4m_\pi^4 w^{-2} \bigg) 
                 -\, 32 g_A^2\, \bigg( m_\pi^2+{7\over 16}q^2 \bigg) 
                 + w^2 \, \Bigg] \,, 
\\
V_{LS} &=& {g_A^4 \, L(q) \over 4\pi^2 M_N^2 f_\pi^4} 
\bigg( {11 \over 32} q^2 +m_\pi^4 w^{-2}\bigg) \,,
\\
W_{LS} &=&  { L(q) \over 256 \pi^2 M_N^2 f_\pi^4}
\Bigg[\, 16 g_A^2\, \bigg( m_\pi^2+{3\over 8}q^2\bigg)
%\nonumber \\ && 
+ \,\frac43\, g_A^4\, \bigg( 4m_\pi^4 w^{-2}-{11\over 4}q^2 -9m_\pi^2 \bigg) 
                      -w^2 \, \Bigg] \,,
\\
V_{\sigma L} &=& {g_A^4 \, L(q) \over 32\pi^2 M_N^2 f_\pi^4}\;.
\label{eq_4M2sL}
\end{eqnarray} 
In the above expressions, we have replaced the $p^2$ dependence used in
Ref.~\cite{Kai01b} by a $k^2$ dependence applying the (on-shell)
identity
\begin{equation}
p^2=\frac14 q^2 + k^2 \,.
\end{equation}

\subsubsection{Two-loop contributions}

The two-loop contributions are quite involved. 
In Fig.~\ref{fig_diag3}, we attempt a graphical representation of
this class. The gray disk stands for all one-loop $\pi N$ graphs
which are shown in some detail in the lower part of the figure.
Not all of the numerous graphs are displayed. Some of the missing ones
are obtained by permutation of the vertices along the nucleon line,
others by inverting initial and final states.
Vertices denoted by a small dot are from the
leading order $\pi N$ Lagrangian
$\widehat{\cal L}^{(1)}_{\pi N}$,
Eq.~(\protect\ref{eq_L1}), except
for the  $4\pi$ vertices which are from 
${\cal L}^{(2)}_{\pi\pi}$, Eq.~(\ref{eq_pipi}).
The solid square represents vertices proportional to the LECs
$d_i$ which are introduced by the third order Lagrangian 
${\cal L}^{(3)}_{\pi N}$,
Eq.~(\ref{eq_L3rel}).
The $d_i$ vertices occur actually in one-loop $NN$ diagrams, but
we list them among the two-loop $NN$ contributions because they
are needed to absorb divergences generated by one-loop $\pi N$ graphs.
Using techniques from dispersion theory,
Kaiser~\cite{Kai01a} calculated the imaginary parts of the
$NN$ amplitudes, Im $V_\alpha(i\mu)$ and Im $W_\alpha(i\mu)$,
which result from analytic continuation to time-like
momentum transfer $q=i\mu-0^+$ with $\mu\geq 2m_\pi$.
We will first state these expressions and, then, further elaborate
on them.
\begin{eqnarray} 
{\rm Im}\, V_C(i\mu) &=& 
-{3g_A^4 (\mu^2-2m_\pi^2) \over \pi \mu (4f_\pi)^6} 
\left[ (m_\pi^2-2\mu^2) \left( 2m_\pi +{2m_\pi^2 -\mu^2 \over
2\mu} \ln{\mu+2m_\pi \over \mu-2m_\pi} \right) 
%\right.  \nonumber \\ && \left.
+4g_A^2 m_\pi(2m_\pi^2-\mu^2) \right] \,,
\\
{\rm Im}\, W_C(i\mu)&=& 
{\rm Im}\, W_C^{(a)}(i\mu) +  {\rm Im}\, W_C^{(b)}(i\mu) 
\\
\mbox{\rm with\hspace*{1.3cm}}
\nonumber \\
{\rm Im}\, W_C^{(a)}(i\mu)&=& 
-{2\kappa \over 3\mu (8\pi f_\pi^2)^3} 
\int_0^1 dx\, 
\Big[ g_A^2(2m_\pi^2-\mu^2) +2(g_A^2-1)\kappa^2x^2 \Big]
\nonumber \\
&& \times \bigg\{ 
\,96 \pi^2 f_\pi^2 
\left[ (2m_\pi^2-\mu^2)(\bar{d}_1 +\bar{d}_2) 
-2\kappa^2x^2 \bar{d}_3
+4m_\pi^2 \bar{d}_5 \right] 
\nonumber \\
&& 
+\left[ 4m_\pi^2 (1+2g_A^2) -\mu^2(1+5g_A^2)\right] 
{\kappa\over \mu} \ln {\mu +2\kappa\over 2m_\pi} \,
%\right.  \nonumber \\ && \left.
+\,{\mu^2 \over 12} (5+13g_A^2) -2m_\pi^2 (1+2g_A^2) 
\bigg\}   
\\
\mbox{\rm and\hspace*{1.4cm}}
\nonumber \\
{\rm Im}\, W_C^{(b)}(i\mu)&=& 
-{2\kappa \over 3\mu (8\pi f_\pi^2)^3} 
\int_0^1 dx\, 
\Big[ g_A^2(2m_\pi^2-\mu^2) +2(g_A^2-1)\kappa^2x^2 \Big]
\nonumber \\
&& \times \left\{ 
-\,3\kappa^2x^2 
+6 \kappa x \sqrt{m_\pi^2 +\kappa^2 x^2} \ln{ \kappa x +\sqrt{m_\pi^2 +\kappa^2 x^2}\over  m_\pi}
\right.
\nonumber \\
&&
\left.
+g_A^4\left(\mu^2 -2\kappa^2 x^2 -2m_\pi^2\right) 
%\right.  \nonumber \\ && \left.  \times 
\left[ {5\over 6} +{m_\pi^2\over \kappa^2 x^2} 
-\left( 1 +{m_\pi^2\over \kappa^2 x^2} \right)^{3/2} 
\ln{ \kappa x +\sqrt{m_\pi^2 +\kappa^2 x^2}\over  m_\pi} \right] 
\right\}\,,   
\\
{\rm Im}\, V_S(i\mu) &=& {\rm Im}\, V_S^{(a)}(i\mu) + {\rm Im}\, V_S^{(b)}(i\mu) 
=\mu^2\,{\rm Im}\, V_T(i\mu)  
=\mu^2\,{\rm Im}\, V_T^{(a)}(i\mu)  +\mu^2\,{\rm Im}\, V_T^{(b)}(i\mu)  
\\
\mbox{\rm with\hspace*{1.3cm}}
\nonumber \\ 
{\rm Im}\, V_S^{(a)}(i\mu) &=&\mu^2\,{\rm Im}\, V_T^{(a)}(i\mu) = 
-{3g_A^2\mu \kappa^3 \over 16\pi f_\pi^4} 
\int_0^1 dx(1-x^2)
\left(\bar{d}_{14}-\bar{d}_{15}\right) 
\\
\mbox{\rm and\hspace*{1.4cm}}
\nonumber \\ 
{\rm Im}\, V_S^{(b)}(i\mu) &=&\mu^2\,{\rm Im}\, V_T^{(b)}(i\mu) = 
-{2g_A^6\mu \kappa^3 \over (8\pi f_\pi^2)^3} 
\int_0^1 dx(1-x^2)
\left[ 
-{1\over 6}+{m_\pi^2 \over \kappa^2x^2} 
%\right.  \nonumber \\ && \left.  \hspace*{3.0cm}
-\left( 1+{m_\pi^2 \over \kappa^2x^2} \right)^{3/2} 
\ln{ \kappa x +\sqrt{m_\pi^2 +\kappa^2 x^2}\over  m_\pi}
\right]
\,, \\ 
{\rm Im}\, W_S(i\mu) &=& \mu^2 \,{\rm Im}\, W_T(i\mu) = 
-{g_A^4(\mu^2-4m_\pi^2) \over \pi (4f_\pi)^6} 
\left[ \left( m_\pi^2 -{\mu^2 \over 4} \right)
\ln{\mu+2m_\pi \over \mu-2m_\pi} 
%\right.  \nonumber \\ && \left.
+(1+2g_A^2)\mu  m_\pi\right]
\,, \\
\end{eqnarray}
where $\kappa \equiv \sqrt{\mu^2/4-m_\pi^2}$.

We need the momentum space amplitudes $V_\alpha(q)$ and $W_\alpha(q)$
which can be obtained from
the above expressions by means of the dispersion integrals:
\begin{eqnarray} 
V_{C,S}(q) &=& 
-{2 q^6 \over \pi} \int_{2m_\pi}^\infty d\mu \,
{{\rm Im\,}V_{C,S}(i \mu) \over \mu^5 (\mu^2+q^2) }\,, 
\\
V_T(q) &=& 
{2 q^4 \over \pi} \int_{2m_\pi}^\infty d\mu \,
{{\rm Im\,}V_T(i \mu) \over \mu^3 (\mu^2+q^2) }\,, 
\end{eqnarray}
and similarly for $W_{C,S,T}$.

We have evaluated these dispersion integrals and obtain:
\begin{eqnarray}
V_C (q) & = &
 \frac{3g_A^4 \widetilde{w}^2 A(q)}{1024 \pi^2 f_\pi^6}
\left[ ( m_\pi^2 + 2q^2 )
\left( 2m_\pi + \widetilde{w}^2 A(q) \right)
+ 4g_A^2 m_\pi \widetilde{w}^2 \right]
\,,
\label{eq_42lC}
\\
W_C(q) & = & W_C^{(a)}(q) + W_C^{(b)}(q)
\\
\mbox{\rm with\hspace*{1cm}}
\nonumber \\
W_C^{(a)}(q) & = &
\frac{L(q)}{18432 \pi^4 f_\pi^6}
\Bigg\{
192 \pi^2 f_\pi^2
w^2 \bar{d}_3
\left[2g_A^2\widetilde{w}^2-\frac35(g_A^2-1)w^2\right]
\nonumber \\
&&
+\left[6g_A^2\widetilde{w}^2-(g_A^2-1)w^2\right]
\Bigg[
 384\pi^2f_\pi^2
\left(\widetilde{w}^2(\bar{d}_1+\bar{d}_2)+4m_\pi^2\bar{d}_5\right)
\nonumber \\
&&
+L(q)
\left(4m_\pi^2(1+2g_A^2)+q^2(1+5g_A^2)\right)
-\left(\frac{q^2}{3}(5+13g_A^2)+8m_\pi^2(1+2g_A^2)\right)
\Bigg]
\Bigg\}
\label{eq_WC}
\\
\mbox{\rm and\hspace*{1.1cm}}
\nonumber \\
W_C^{(b)}(q) & = &
-{2 q^6 \over \pi} \int_{2m_\pi}^\infty d\mu \,
{{\rm Im\,}W_C^{(b)}(i \mu) \over \mu^5 (\mu^2+q^2) }\,, 
\\
V_T(q)  &=& V_T^{(a)}(q) + V_T^{(b)}(q) =
 - {1 \over q^2}V_S(q) 
= - {1\over q^2}\left(V_S^{(a)}(q)+V_S^{(b)}(q)\right)
\\
\mbox{\rm with\hspace*{1cm}}
\nonumber \\
V_T^{(a)}(q)  &=& - {1\over q^2}V_S^{(a)}(q) =
-\frac{g_A^2 w^2 L(q)}{32 \pi^2 f_\pi^4}
(\bar{d}_{14} - \bar{d}_{15}) 
\label{eq_VT}
\\
\mbox{\rm and\hspace*{1.1cm}}
\nonumber \\
V_T^{(b)}(q)  &=& - {1\over q^2}V_S^{(b)}(q) =
{2 q^4 \over \pi} \int_{2m_\pi}^\infty d\mu \,
{{\rm Im\,}V_T^{(b)}(i \mu) \over \mu^3 (\mu^2+q^2) }\,, 
\\
W_T(q) &=& - {1 \over q^2}W_S(q) =
\frac{g_A^4 w^2 A(q)}{2048 \pi^2 f_\pi^6}
\left[ w^2 A(q) + 2m_\pi (1 + 2 g_A^2) \right]
\label{eq_42lT}
\end{eqnarray}

We were able to find analytic solutions for all dispersion
integrals except 
$W_C^{(b)}$ 
and 
$V_T^{(b)}$ 
(and 
$V_S^{(b)}$).
The analytic solutions hold modulo polynomials.
We have checked the importance of those contributions where the
integrations have to be performed numerically. It turns out
that the combined effect on $NN$ phase shifts from
$W_C^{(b)}$, $V_T^{(b)}$, and $V_S^{(b)}$
is smaller than 0.1 deg in $F$ and $G$ waves
and smaller than 0.01 deg in $H$ waves,
at $T_{\rm lab} = 300$ MeV (and less at lower energies). 
This renders these contributions negligible, a fact that may be of
interest in future chiral $NN$ potential developments
where computing time could be an issue.
We stress, however, that in all phase shift calculations of this paper
(presented in Sec.~IV, below)
the contributions from
$W_C^{(b)}$, $V_T^{(b)}$, and $V_S^{(b)}$
are always included in all fourth order results.

In Eqs.~(\ref{eq_WC}) and (\ref{eq_VT}), we use the scale-independent
LECs, $\bar{d}_i$, which are obtained by combining the 
scale-dependent ones, $d_i^r (\lambda)$, 
with the chiral logarithmus,
$\ln (m_\pi/\lambda)$, or equivalently $\bar{d}_i = d^r_i(m_\pi)$.
The scale-dependent LECs, $d_i^r (\lambda)$, 
are a consequence of renormalization.
For more details about this issue, see Ref.~\cite{FMS98}.

\section{$NN$ scattering in peripheral partial waves}

In this section, we will calculate the phase shifts
that result from the $NN$ amplitudes presented in the previous
section and compare them to the empirical phase shifts
as well as to the predictions from conventional meson theory. 
For this comparison to be realistic, we must also include
the one-pion-exchange (OPE) amplitude and the iterated
one-pion-exchange, which we will explain first.
We then describe in detail how the phase shifts
are calculated. Finally, we show phase parameters for $F$ and higher
partial waves and energies below 300 MeV.

\subsection{OPE and iterated OPE}

Throughout this paper, we consider neutron-proton ($np$) scattering 
and take the charge-dependence of OPE due to pion-mass splitting
into account, since it is appreciable. Introducing the definition,
\begin{equation}
V_\pi (m_\pi) \equiv -\, 
\frac{1}{(2\pi)^3} 
\frac{g_A^2}{4f_\pi^2}\, \frac{
\vec \sigma_1 \cdot \vec q \,\, \vec \sigma_2 \cdot \vec q}
{q^2 + m_\pi^2} 
\,,
\end{equation}
the charge-dependent OPE for $np$ scattering is given by,
\begin{equation}
V_{\rm OPE} ({\vec p}~', \vec p) 
= -V_\pi(m_{\pi^0}) + (-1)^{I+1}\, 2\, V_\pi (m_{\pi^\pm})
\,,
\label{eq_OPE}
\end{equation}
where $I$ denotes the isospin of the two-nucleon system.
We use $m_{\pi^0}=134.9766$ MeV and
 $m_{\pi^\pm}=139.5702$ MeV~\cite{PDG00}.

The twice iterated OPE generates the iterative part of the
$2\pi$-exchange, which is
\begin{equation}
V_{2\pi, it} ({\vec p}~',{\vec p})  = 
\frac{M_N^2}{E_p} \int d^3p''\:
\frac{V_{\rm OPE} ({\vec p}~',{\vec p}~'')\,
V_{\rm OPE} ({\vec p}~'',{\vec p})} 
{{ p}^{2}-{p''}^{2}+i\epsilon}
\,,
\label{eq_2piit}
\end{equation}
where, for $M_N$, we use twice the reduced
mass of proton and neutron,
\begin{equation}
M_N  =  \frac{2M_pM_n}{M_p+M_n} = 938.9182 \mbox{ MeV}
\,,
\end{equation}
and $E_p\equiv \sqrt{M^2_N+p^2}$.

The $T$-matrix considered in this study is, 
\begin{equation}
T({\vec p}~',\vec p) = V_{\rm OPE} ({\vec p}~', \vec p) 
+V_{2\pi, it} ({\vec p}~',{\vec p}) +V_{2\pi, irr} 
({\vec p}~',{\vec p}) 
\label{eq_tall}
\end{equation}
where $V_{2\pi,irr}$ refers to any or all
of the contributions presented in Sec.~III. 
In the calculation of the latter contributions,
we use the average pion mass $m_\pi = 138.039$ MeV and, thus,
neglect the charge-dependence due to pion-mass splitting.
The charge-dependence that emerges from irreducible $2\pi$ 
exchange was investigated in Ref.~\cite{LM98} and found to be
negligible for partial waves with $L\geq 3$.

\subsection{Calculating phase shifts}
We perform a partial-wave 
decomposition of the amplitude using
the formalism of Refs.~\cite{JW59,GM61,EAH71}.
For this purpose, we first represent 
$T({\vec p}~',\vec p)$, Eq.~(\ref{eq_tall}), in terms of helicity states yielding
$\langle {\vec p}~'\lambda_1'\lambda_2'|T|{\vec p}\, \lambda_1\lambda_2\rangle$.
Note that the helicity $\lambda_i$ of particle $i$ (with $i=1$ or 2) 
is the eigenvalue of the helicity operator 
$\frac12 {\vec  \sigma}_i \cdot {\vec p}_i/|{\vec p}_i|$ which is 
$\pm \frac12$. 
Decomposition into angular momentum states is accomplished by
\begin{eqnarray}
\langle \lambda_1' \lambda_2' |T^J(p',p)| \lambda_1 \lambda_2 \rangle =
2\pi \int^{+1}_{-1} d(\cos \theta)\: d^J_{\lambda_1-\lambda_2, 
\lambda_1'-\lambda_2'}(\theta)
\langle {\vec p}~'\lambda_1'\lambda_2'|T|{\vec p}\, \lambda_1\lambda_2\rangle
\label{eq_vhel}
\end{eqnarray}
where $\theta$ is the angle between ${\vec p}~'$ and ${\vec p}$ and
$d^J_{m,m'}(\theta)$ are the conventional reduced rotation matrices
which can be expressed in terms of Legendre 
polynominals $P_J(\cos \theta)$. 
Time-reversal invariance, parity conservation, and spin conservation
(which is a consequence of isospin conservation and the Pauli principle)
imply that 
only five of the 16 helicity amplitudes are independent. 
For the five amplitudes, we 
choose the following set:
\begin{eqnarray}
{T}^J_1(p,p)&\equiv& \langle ++|{T}^J(p,p)|++ \rangle \nonumber \\
{T}^J_2(p,p)&\equiv& \langle ++|{T}^J(p,p)|-- \rangle \nonumber \\
{T}^J_3(p,p)&\equiv& \langle +-|{T}^J(p,p)|+- \rangle \nonumber \\
{T}^J_4(p,p)&\equiv& \langle +-|{T}^J(p,p)|-+ \rangle \nonumber \\
{T}^J_5(p,p)&\equiv& \langle ++|{T}^J(p,p)|+- \rangle
\label{eq_transf}
\end{eqnarray}
where $\pm$ stands for $\pm \frac12$, and where the repeated argument $(p,p)$
stresses the fact that our consideration is restricted to the on-shell
amplitude.
The following linear combinations of helicity amplitudes will prove
to be useful:
\begin{eqnarray}
^0{T}^J&\equiv&{T}^J_1 - {T}^J_2  \nonumber \\
^1{T}^J&\equiv&{T}^J_3 - {T}^J_4  \nonumber \\
^{12}{T}^J&\equiv&{T}^J_1 + {T}^J_2 \nonumber \\
^{34}{T}^J&\equiv&{T}^J_3 + {T}^J_4 \nonumber \\
^{55}{T}^J&\equiv&2{T}^J_5 
\label{eq_lincom}
\end{eqnarray}
More common in nuclear physics is the representation of 
two-nucleon states
in terms of an 
$|LSJM\rangle$ basis, 
where $S$ denotes the total spin, $L$ the total orbital 
angular momentum, and $J$ the total angular momentum with 
projection $M$. 
In this basis, we will denote the ${T}$ matrix elements by
${T}^{JS}_{L',L}\equiv \langle L'SJM|{T}|LSJM\rangle$.
  These are obtained from the helicity state matrix 
elements by the following unitary transformation:\\
{ Spin singlet}
\begin{equation}
{T}^{J0}_{J,J}\: =\: ^0{T}^J
\: .
\end{equation}
{ Uncoupled spin triplet} 
\begin{equation}
{T}^{J1}_{J,J}\: =\: ^1{T}^J
\: .
\end{equation}
{ Coupled triplet states}
\begin{eqnarray}
{T}^{J1}_{J-1,J-1} & = & \frac{1}{2J+1} \left[J\: ^{12}{T}^J 
+ (J+1)\: ^{34}{T}^J
+ 2 \sqrt{J(J+1)} \: ^{55}{T}^J \right]
\nonumber \\
{T}^{J1}_{J+1,J+1} & = & \frac{1}{2J+1} \left[(J+1)\: ^{12}{T}^J 
+ J\: ^{34}{T}^J
- 2 \sqrt{J(J+1)} \: ^{55}{T}^J \right]
\nonumber \\
{T}^{J1}_{J-1,J+1} 
& = & \frac{1}{2J+1} \left[\sqrt{J(J+1)}
(\: ^{12}{T}^J -\: ^{34}{T}^J)
+\: ^{55}{T}^J \right]
\nonumber  \\
{T}^{J1}_{J+1,J-1} 
& = & 
{T}^{J1}_{J-1,J+1} 
\: .
\label{eq_lsj}
\end{eqnarray}
The matrix elements for the five spin-dependent operators involved in
Eq.~(\ref{eq_nnamp}) in a helicity state basis, Eqs.~(\ref{eq_vhel}),
as well as in $|LSJM\rangle$ basis, Eq.~(\ref{eq_lsj}),
are given in section 4 of Ref.~\cite{EAH71}.
Note that, for the amplitudes
${T}^{J1}_{J-1,J+1}$ and ${T}^{J1}_{J+1,J-1}$,
we use a sign convention that differs by a factor $(-1)$ from
the one used in Ref.~\cite{EAH71}. 

We consider neutron-proton scattering and determine the
CM on-shell nucleon momentum $p$ using correct
relativistic kinematics:
\begin{equation}
p^2  =  \frac{M_p^2 T_{lab} (T_{lab} + 2M_n)}
               {(M_p + M_n)^2 + 2T_{lab} M_p} \: , 
\label{eq_relkin}
\end{equation}
where $M_p=938.2720$ MeV is the proton mass, 
$M_n=939.5653$ MeV the neutron mass~\cite{PDG00}, 
and $T_{lab}$ is the kinetic energy of
the incident nucleon in the laboratory system.

The on-shell $S$-matrix is related to the on-shell $T$-matrix
by
\begin{equation}
S^{JS}_{L'L} (T_{lab}) 
= \delta_{L'L} + 2i \,
\tau^{JS}_{L'L} (p,p) 
\,, 
\label{eq_tmat}
\end{equation}
with
\begin{equation}
\tau^{JS}_{L'L} (p,p) 
\equiv 
-\frac{\pi}{2} \frac{M^2_N}{E_p}\, p \,
T^{JS}_{L'L} (p,p) 
\,. 
\end{equation}

For an uncoupled partial wave,
the phase shifts $\delta^{JS}_J (T_{lab})$
parametrizes the partial-wave $S$-matrix in the form
\begin{equation}
S^{JS}_{JJ} (T_{lab}) 
= \eta^{JS}_J(T_{lab})\, 
e^{2i \, \delta^{JS}_J (T_{lab})}
\,, 
\end{equation}
implying
\begin{equation}
\tan 2 \delta^{JS}_J (T_{lab})
= \frac{2\, {\rm Re}\, \tau^{JS}_{JJ} (p,p)}
{1 - 2\, {\rm Im}\, \tau^{JS}_{JJ} (p,p)} 
\, .
\end{equation}
The real parameter $\eta^{JS}_J(T_{lab})$, which is given by
\begin{equation}
\eta^{JS}_J(T_{lab}) = \left| S^{JS}_{JJ} (T_{lab}) \right|
\,,
\end{equation}
tells us to what extend unitarity is observed (ideally, it
should be unity).

For coupled partial waves, we use the parametrization introduced
by Stapp {\it et al.}~\cite{SYM57} 
(commonly known as `bar' phase shifts, {\it but we denote them
simply by $\delta^J_\pm$ and $\epsilon_J$}), 
\begin{equation}
\left(
\begin{array}{cc}
S^J_{--} & S^J_{-+} \\
S^J_{+-} & S^J_{++}
\end{array}
\right)
=
\left(
\begin{array}{cc}
(\eta^J_-)^{\frac12}
e^{i\delta^J_-} 
& 0 \\
0 & 
(\eta^J_+)^{\frac12}
e^{i\delta^J_+}
\end{array}
\right)
\left(
\begin{array}{cc}
\cos 2\epsilon_J & i\sin 2\epsilon_J\\
i\sin 2\epsilon_J & \cos 2\epsilon_J 
\end{array}
\right)
\left(
\begin{array}{cc}
(\eta^J_-)^{\frac12}
e^{i\delta^J_-} 
& 0 \\
0 & 
(\eta^J_+)^{\frac12}
e^{i\delta^J_+}
\end{array}
\right)
\,,
\end{equation}
where the subscript `+' stands for `$J+1$' and `$-$' for `$J-1$'
and where the superscript $S=1$ as well as the argument $T_{lab}$ are suppressed.
The explicit formulae for the resulting phase parameters are,
\begin{eqnarray}
\tan 2\delta^J_\pm & = & \frac{{\rm Im}\, 
\left(S^J_{\pm\pm}/\cos 2\epsilon_J\right)}
{{\rm Re}\, \left(S^J_{\pm\pm}/\cos 2\epsilon_J\right)}
\,,
\\
\tan 2\epsilon_J & = & \frac{-i\,S^J_{+-}}
{\sqrt{S^J_{++}S^J_{--}}}
\,,
\\
\eta^J_\pm & = & \left|\frac{S^J_{\pm\pm}}{\cos 2\epsilon_J} \right|
\,.
\label{eq_eta}
\end{eqnarray}
The parameters $\delta^J_{\pm}$ and $\eta^J_\pm$ are always real,
while the mixing parameter $\epsilon_J$ is real if $\eta^J_\pm=1$
and complex otherwise.

We note that since the $T$-matrix is calculated perturbatively 
[cf.\ Eq.~(\ref{eq_tall})], unitarity is (slightly) violated. 
Through the parameter $\eta^{JS}_L$,
the above formalism provides precise information on
the violation of unitarity.
It turns out that for the cases considered in this paper
(namely partial waves with $L\geq 3$ and $T_{lab}\leq 300$ MeV)
the violation of unitarity is, generally, in the order of 1\% or less.

There exists an alternative method of calculating phase shifts for
which unitarity is perfectly observed.
In this method---known as the $K$-matrix approach---one 
identifies the real part of the amplitude $V$ with  the $K$-matrix. 
For an uncoupled partial-wave, the $S$-matrix element, $S_L$, is defined in terms
of the (real) $K$-matrix element, $\kappa_L$, by
\begin{equation}
S_L (T_{lab}) = \frac{1+i\kappa_L(p,p)}{1-i\kappa_L(p,p)}
\label{eq_kmat}
\end{equation}
which guarantees perfect unitarity and yields the phase shift
\begin{equation}
\tan\: \delta_L(T_{lab}) = \kappa_L(p,p) = -\frac{\pi}{2} {M_N^2 \over E_p}\: p \: K_L(p,p)
\, ,
\end{equation}
with $K_L(p,p)={\rm Re}\, V_L(p,p)$.
Combining Eqs.~(\ref{eq_tmat}) and (\ref{eq_kmat}),
one can write down the $T$-matrix element, $\tau_L$, that is
equivalent to a given $K$-matrix element, $\kappa_L$,
\begin{equation}
\tau_L(p,p) = \frac{\kappa_L(p,p)+i
\kappa_L^2(p,p)}
{1+
\kappa_L^2(p,p)}
\,.
\end{equation}
Obviously, this $T$-matrix includes higher orders of $K$ (and, thus, of $V$)
such that consistent power counting is destroyed.

The bottom line is that
there is no perfect way of calculating phase shifts for
a perturbative amplitude. Either one includes contributions strictly
to a certain order, but violates unitarity, or one
satisfies unitarity, but includes implicitly contributions beyond
the intended order.
To obtain an idea of what uncertainty this dilemma creates, 
we have calculated all phase
shifts presented below both ways: using the $T$-matrix and 
$K$-matrix approach. We found that
the difference between the phase shifts due to the
two different methods 
is smaller than 0.1 deg in $F$ and $G$ waves
and smaller than 0.01 deg in $H$ waves,
at $T_{\rm lab} = 300$ MeV (and less at lower energies). 
Because of this
small difference, we have confidence in our phase shift
calculations. All results presented below have been obtained
using the $T$-matrix approach, Eqs.~(\ref{eq_tmat})-(\ref{eq_eta}).

\subsection{Results}
For the $T$-matrix given in Eq.~(\ref{eq_tall}),
we calculate phase shifts for
partial waves with $L\geq 3$ and $T_{lab}\leq 300$ MeV.
At order four in small momenta, partial waves
with $L\geq 3$ do not receive any contributions from
contact interactions and, thus, the non-polynomial
pion contributions uniquely predict the $F$ and
higher partial waves.
The parameters used in our calculations are shown in Table~I.
In general, we use average masses for nucleon and pion, $M_N$ and $m_\pi$, 
as given in Table~I. There are, however, two exceptions from
this rule. For the evaluation of the CM on-shell momentum, $p$,
we apply correct relativistic kinematics, Eq.~(\ref{eq_relkin}),
which involves the correct and precise values for the
proton and neutron masses. 
For OPE, we use the charge-dependent
expression, Eq.~(\ref{eq_OPE}), which employes the correct and precise
values for the charged and neutral pion masses.

Many determinations of the LECs, $c_i$ and $\bar{d}_i$,
can be found in the literature.
The most reliable way to determine the LECs from empirical
$\pi N$ information is to extract them from the $\pi N$ amplitude
inside the Mandelstam triangle (unphysical region) which can
be constructed with the help of dispersion relations from empirical
$\pi N$ data. This method was used by B\"uttiker and Mei\ss ner~\cite{BM00}.
Unfortunately, the values for $c_2$ and all $\bar{d}_i$ parameters
obtained in Ref.~\cite{BM00} carry uncertainties,
so large that the values are useless.
Therefore, in Table~I, only  $c_1$, $c_3$, and $c_4$
are from Ref.~\cite{BM00}, while the other LECs
are taken from Ref.~\cite{FMS98} where the $\pi N$ amplitude in the
physical region was considered.
To establish a link between $\pi N$ and $NN$, we apply 
the values from the above determinations in our $NN$ calculations.
In general, we use the mean values;
the only exception is $c_3$, where we choose
a value that is, in terms of magnitude, about one standard
deviation below the one from Ref.~\cite{BM00}.
With the exception of $c_3$,
our results do not depend sensitively on
variations of the LECs within the quoted uncertainties.

In Figs.~\ref{fig_f}-\ref{fig_h}, 
we show the phase-shift
predictions for neutron-proton scattering in $F$, $G$, and
$H$ waves for laboratory kinetic energies below 300 MeV.
The orders displayed are defined as follows:
\begin{itemize}
\item
Leading order (LO) is just OPE, Eq.~(\ref{eq_OPE}).
\item
Next-to-leading order (NLO) is OPE plus iterated OPE,
Eq.~(\ref{eq_2piit}), plus the contributions of Sec.~III.A
(order two), Eqs.~(\ref{eq_2C}) and (\ref{eq_2T}).
\item
Next-to-next-to-leading order (denoted by N2LO in the figures)
consists of NLO plus the contributions of Sec.~III.B (order three),
Eqs.~(\ref{eq_3C})-(\ref{eq_3LS}).
\item
Next-to-next-to-next-to-leading order (denoted by N3LO in the figures)
consists of N2LO plus the contributions of Sec.~III.C (order four),
Eqs.~(\ref{eq_4c2C}), 
(\ref{eq_4c2T}), 
(\ref{eq_4cMC})-(\ref{eq_4cMLS}), 
(\ref{eq_4M2C})-(\ref{eq_4M2sL}), 
and (\ref{eq_42lC})-(\ref{eq_42lT}). 
To this order, the phase shifts have never been calculated before.
\end{itemize}
It is clearly seen in 
Figs.~\ref{fig_f}-\ref{fig_h} 
that the leading order $2\pi$ exchange (NLO)
is a rather small contribution, insufficient to explain
the empirical facts. In contrast, the next order (N2LO)
is very large, several times NLO. This is due to the
$\pi\pi N N$ contact interactions proportional
to the LECs $c_i$ that are introduced by the
second order Lagrangian 
${\cal L}^{(2)}_{\pi N}$, Eq.~(\ref{eq_L2rel}). 
These contacts are supposed to simulate the contributions
from intermediate $\Delta$-isobars and correlated $2\pi$
exchange which are known
to be large (see, e.~g., Ref.~\cite{MHE87}). 

All past calculations of $NN$ phase shifts in peripheral partial
waves stopped at order N2LO (or lower). This was very
unsatisfactory, since to this order there is no indication 
that the chiral expansion will ever converge.
The novelty of the present work is the calculation of phase shifts
to N3LO (the details of which are shown in Appendix A). 
Comparison with N2LO reveals that at N3LO a clearly identifiable
trend towards convergence emerges
(Figs.~\ref{fig_f}-\ref{fig_h}). 
In $G$ (except for $^3G_5$, a problem that
is discussed in Appendix A) and $H$ waves, N3LO differs very little from
N2LO, implying that we have reached convergence.
Also $^1F_3$ and $^3F_4$ appear fully converged.
In $^3F_2$ and $^3F_3$, N3LO differs
noticeably from N2LO, but the difference is much smaller than the
one between N2LO and NLO. This is what we perceive as a trend towards
convergence.

In Figs.~\ref{fig_ff}-\ref{fig_hh}, 
we conduct a comparison between the predictions from chiral one- and
two-pion exchange at N3LO and the corresponding predictions from conventional
meson theory (curve `Bonn'). 
As representative for conventional meson theory, we choose the Bonn
meson-exchange model for the $NN$ interaction~\cite{MHE87},
since it contains a comprehensive and thoughtfully constructed
model for $2\pi$ exchange.
This $2\pi$ model includes box and crossed box diagrams with
$NN$, $N\Delta$, and $\Delta\Delta$ intermediate states
as well as direct $\pi\pi$ interaction in $S$- and $P$-waves
(of the $\pi\pi$ system)
consistent with empirical information from $\pi N$ and $\pi\pi$
scattering.
Besides this the Bonn model also includes (repulsive) $\omega$-meson
exchange and irreducible diagrams of $\pi$ and $\rho$ exchange
(which are also repulsive).
In the phase shift predictions displayed in Figs.~\ref{fig_ff}-\ref{fig_hh}, 
the Bonn calculation includes only the OPE and $2\pi$ contributions
from the Bonn model; the short-range contributions are
left out to be consistent with the chiral calculation.
In all waves shown (with the usual exception of $^3G_5$),
we see, in general, good agreement between N3LO and Bonn~\cite{foot3}.
In $^3F_2$ and $^3F_3$ above 150 MeV and in
$^3F_4$ above 250 MeV the chiral model to N3LO is more
attractive than the Bonn $2\pi$ model. Note, however,
that the Bonn model is relativistic and, thus, includes
relativistic corrections up to infinite orders.
Thus, one may speculate that higher orders in 
chiral perturbation theory ($\chi$PT) may create some
repulsion, moving the Bonn and the chiral predictions even closer
together~\cite{foot2}.

The $2\pi$ exchange contribution to the $NN$ interaction
can also be derived
from {\it empirical} $\pi N$ and $\pi\pi$ input
using dispersion theory, which is based upon unitarity,
causality (analyticity), and crossing symmetry.
The amplitude $N\bar{N}\rightarrow \pi\pi$ is
constructed from $\pi N \rightarrow \pi N$
and $\pi N \rightarrow \pi\pi N$ data using 
crossing properties and analytic
continuation; this amplitude is then `squared' to yield 
the $N\bar{N}$ amplitude 
which is related to $NN$ by crossing symmetry~\cite{BJ76}.
The Paris group~\cite{Paris} pursued this path and 
calculated $NN$ phase shifts in peripheral partial waves.
Naively, the dispersion-theoretic approach is
the ideal one, since it is based exclusively on empirical
information. Unfortunately, in practice,
quite a few uncertainties enter into the approach.
First, there are ambiguities
in the analytic continuation and, second, 
the dispersion integrals have to be cut off at a certain
momentum to ensure reasonable results.
In Ref.~\cite{MHE87}, a thorough comparison was conducted
between the predictions by the Bonn model and the
Paris approach and it was demonstrated that the Bonn
predictions always lie comfortably within the range of uncertainty
of the dispersion-theoretic results.
Therefore, there is no need to perform a
separate comparison of our chiral N3LO predictions with
dispersion theory, since it would not add anything that we
cannot conclude from Figs.~\ref{fig_ff}-\ref{fig_hh}.

Finally, we like to compare the
predictions with the empirical phase shifts.
In $G$ (except $^3G_5$) and $H$ waves there is excellent
agreement between the N3LO predictions and the data.
On the other hand, in $F$ waves the predictions above 200 MeV are,
in general, too attractive. Note, however, that this is
also true for the predictions by the Bonn $\pi + 2\pi$
model. In the full Bonn model, also (repulsive)
$\omega$ and $\pi\rho$ exchanges are included which bring
the predictions to agreement with the data.
The exchange of a $\omega$ meson or  combined $\pi\rho$
exchange are $3\pi$ exchanges.
Three-pion exchange occurs first at chiral order four.
It has be investigated by Kaiser~\cite{Kai99} and found to be
totally negligible, at this order.
However, $3\pi$ exchange at order five appears to be sizable~\cite{Kai01}
and may have impact on $F$ waves.
Besides this, there is the usual short-range phenomenology.
In $\chi$PT, this short-range interaction
is parametrized in terms of four-nucleon contact terms
(since heavy mesons do not have a place in that theory).
Contact terms of order six are effective in $F$-waves.
In summary, the remaining small 
discrepancies between the N3LO predictions and the empirical
phase shifts may be straightened out in
fifth or sixth order of $\chi$PT.

\section{Conclusions and further discussion}

We have calculated the phase shifts for peripheral partial waves
($L\geq 3$) of neutron-proton scattering at order four
(N$^3$LO) in $\chi$PT. The two most important results from
this study are:
\begin{itemize}
\item
At N$^3$LO, the chiral expansion reveals a clearly identifiable
signature of convergence.
\item
There is good agreement between the N$^3$LO prediction and the 
corresponding one from conventional meson theory 
as represented by the Bonn Full Model~\cite{MHE87}.
\end{itemize}
The conclusion from the above two facts is that {\it the chiral
expansion for the $NN$ problem is now under control.}
As a consequence, one can state with confidence that
the $\chi$PT approach to the $NN$ interaction
is a valid one. 

Besides the above fundamentally important statements,
our study has also some more specific implications.
A controversial issue that has recently drawn a lot of
attention~\cite{Epe02} is the question whether the LECs extracted from
$\pi N$ are consistent with $NN$.
After discussing dispersion theory in the previous section,
one may wonder how this can be an issue in the year of 2002.
In the early 1970's. 
the Stony Brook~\cite{CDR72,BJ76} and the Paris~\cite{Vin73,Paris} groups showed
independently that $\pi N$ and $NN$ are consistent,
based upon dispersion-theoretic calculations.
Since dispersion theory is a model-independent approach,
the finding is of general validity. 
Therfore, if 30 years later a specific theory has problems with
the consistency of $\pi N$ and $NN$, then that theory can only be wrong.
Fortunately, we can confirm that $\chi$PT for $\pi N$ and
$NN$ does yield consistent results, as we will explain now in 
more detail.

The reliable way to investigate this issue is to
use an approach that does not contain any parameters
except for the LECs. This is exactly true for our calculations
since we do not use any cutoffs and calculate the $T$ matrix
directly up to a well defined order.
We then vary the LECs within their one-standard deviation range
from the $\pi N$ determinations (cf.\ Table~I).
We find that these variations do not create any essential changes of
the predicted peripheral $NN$ phase shifts shown in 
Figs.~\ref{fig_f}-\ref{fig_hh}, except for $c_3$.
Thus, the focus is on $c_3$.
We find that $c_3 = -3.4$ GeV$^{-1}$ is consistent with the empirical
phase shifts as well as the results from dispersion theory and 
conventional meson theory as demonstrated in
Figs.~\ref{fig_ff}-\ref{fig_hh}.
This choice for $c_3$ is within one standard
deviation of its $\pi N$ determination and, thus,
the consistency of $\pi N$ and $NN$ in $\chi$PT
at order four is established.

In view of the transparent and conclusive consideration presented above,
it is highly disturbing to find in the literature
very different values for $c_3$, allegedly based upon $NN$.
In Ref.~\cite{Ren99}, it its claimed that the value
$c_3 =-5.08\pm 0.28$ GeV$^{-1}$ 
emerges from the world $pp$ data
below 350 MeV, whereas Ref.~\cite{Epe02}
asserts that
$c_3 =-1.15$ GeV$^{-1}$ is implied by the $NN$ phase shifts.
The two values differ by more than 400\% which is reason for
deep concern.

In Fig~\ref{fig_3f4}, we show the predictions at
order four for the three values for $c_3$ under debate.
We have chosen the $^3F_4$ as representative example of
a peripheral partial wave since it has a rather large
contribution from $2\pi$ exchange.
Moreover, the LEC $c_4$ is ineffective in $^3F_4$ such that
differences in the choices for $c_4$ do not distort the picture
in this partial wave.
This fact makes $^3F_4$ special for the discussion of $c_3$.

Figure~\ref{fig_3f4} reveals that
the chiral $2\pi$ exchange depends most sensitively on $c_3$.
It is clearly seen that the Nijmegen choice 
$c_3 = -5.08$ GeV$^{-1}$~\cite{Ren99}
leads to too much attraction,
while the value $c_3 = -1.15$ GeV$^{-1}$,
advocated in Ref.~\cite{Epe02},
is far too small (in terms of magnitude) 
since it results in an almost vanishing $2\pi$ exchange 
contribution---quite in contrast to the empirical $NN$ facts, the dispersion
theoretic result, and the Bonn model.

One reason for the difference between the Nijmegen value 
and ours could be that 
their analysis is conducted at N$^2$LO,
while we go to N$^3$LO. 
However, as demonstrated in Figs.~\ref{fig_f}-\ref{fig_h}, 
N$^3$LO is not that different from N$^2$LO and, therefore, 
not the main reason for the difference.
More crucial is the fact that, in the Nijmegen analysis,
the chiral $2\pi$ exchange potential, represented as local
$r$-space function, is cutoff at $r=1.4$ fm (i.~e., it is set
to zero for $r\leq 1.4$ fm)~\cite{foot}. This cutoff suppresses
the $2\pi$ contribution, also, in peripheral waves.
If the $2\pi$ potential is suppressed by phenomenology
then, of course, stronger values for $c_3$ are necessary,
resulting in a highly model-dependent determination of $c_3$.
For example, if we multiply all non-iterative $2\pi$ contributions
by $\exp[-(p^{2n}+p'~^{2n})/\Lambda^{2n}]$ with 
$\Lambda \approx 400$ MeV and $n=2$,
then with $c_3=-5.08$ GeV$^{-1}$ we obtain a good reproduction
of the peripheral partial wave phase shifts.
Note that
$\Lambda \approx 400$ MeV is roughly equivalent to
a $r$-space cutoff of about 0.5 fm, which is not even close to the
cutoff used in the Nijmegen analysis.
In fact, the Nijmegen $r$-space cutoff of $r=1.4$ fm is
equivalent to  a momentum-space cutoff $\Lambda\approx m_\pi$
which is bound to kill the $2\pi$ exchange contribution (which has
a momentum-space range of $2m_\pi$ and larger). 
To revive it, unrealistically large parameters are necessary.

The motivation underlying the value for $c_3$
advocated in Ref.~\cite{Epe02}, is quite different
from the Nijmegen scenario.
In Ref.~\cite{Epe02}, $c_3$ was adjusted to the $D$ waves of $NN$
scattering, which are notoriously too attractive.
With their choice, $c_3=-1.15$, the $D$ waves are, indeed,
about right, whereas the $F$ waves are drastically underpredicted.
This violates an important rule:
{\it The higher the partial, the higher the priority.}
The reason for this rule is that we have more trust
in the long-range contributions to the nuclear force
than in the short-range ones. The $\pi$ + $2\pi$ contributions
to the nuclear force rule the $F$ and higher partial
waves, not the $D$ waves.
If $D$ waves do not come out right, then one can think of plenty of
short-range contributions to fix it.
If $F$ and higher partial waves are wrong, there is
no fix.

In summary, a realistic choice for the important LEC $c_3$
is -3.4 GeV$^{-1}$ and one may deliberately assign an uncertainty
of $\pm 10$\% to this value. Substantially different values are 
unrealistic as clearly demonstrated in
Figure~\ref{fig_3f4}.

\acknowledgements
We like to thank N. Kaiser for substantial advice throughout
this project. Interesting communications with J. W. Durso 
are gratefully acknowledged.
This work was supported in part by the U.S. National Science
Foundation under Grant No.~PHY-0099444 and by the Ram\'on Areces
Foundation (Spain).

\appendix

\section{Details of order-four contributions to peripheral
partial wave phase shifts}

The order four, consists of very many contributions (cf.\ Sec.~III.C
and Figs.~\ref{fig_diag2} and \ref{fig_diag3}).
Here, we show how the various contributions of order four
impact $NN$ phase shifts in peripheral partial waves.
For this purpose, we display in 
Fig.~\ref{fig_1f3} 
phase shifts
for four important peripheral partial waves, namely,
$^1F_3$, $^3F_3$, $^3F_4$, and $^3G_5$.
In each frame, the following individual
order-four contributions are shown:
\begin{itemize}
\item
$c_i^2$ graph, first row of 
Fig.~\ref{fig_diag2},
Eqs.~(\ref{eq_4c2C}) and (\ref{eq_4c2T}),
denoted by `c2' 
in Fig.~\ref{fig_1f3}. 
\item
$c_i/M_N$ contributions (denoted by `c/M'),
second row of Fig.~\ref{fig_diag2},
Eqs.~(\ref{eq_4cMC})-(\ref{eq_4cMLS}). 
\item
$1/M^2_n$ corrections (`1/M2'),
row three to six of
Fig.~\ref{fig_diag2},
Eqs.~(\ref{eq_4M2C})-(\ref{eq_4M2sL}). 
\item
Two-loop contributions without the terms proportional
to $\bar{d}_i$ (`2-L'); 
Fig.~\ref{fig_diag3}, but without the solid square;
Eqs.~(\ref{eq_42lC})-(\ref{eq_42lT}), but with
all $\bar{d}_i \equiv 0$. 
\item
Two-loop contributions including the terms proportional
to $\bar{d}_i$ (denoted by `d' in Fig.~\ref{fig_1f3}); 
Fig.~\ref{fig_diag3};
Eqs.~(\ref{eq_42lC})-(\ref{eq_42lT}) with the
$\bar{d}_i$ parameters as given in Table~I.
\end{itemize}
Starting with the result at N2LO, curve (1),
the individual N3LO contributions are added up
successively in the order given in parenthesis
next to each curve.
The last curve in this series, curve (6),
is the full N3LO result.

The $c_i^2$ graph generates large attraction in all partial
waves [cf.\ differences between curves (1) and (2) in 
Fig.~\ref{fig_1f3}].
This attraction is compensated by repulsion from the
$c_i/M_N$ diagrams, in most partial waves; the exception
is $^1F_3$ where $c_i/M_N$ adds more  attraction [curve (3)].
The $1/M_N^2$ corrections [difference between curves (3)
and (4)] are typically small.
Finally, the two-loop contributions create substantial repulsion
in $^1F_3$ and $^3G_5$ which brings $^1F_3$ into good agreement
with the data while causing a discrepancy for $^3G_5$. 
In $^3F_3$ and $^3F_4$, there are large cancelations
between the `pure' two-loop graphs and the $\bar{d}_i$ terms,
making the net two-loop contribution rather small.

A pivotal role in the above game is played by $W_S$, 
Eq.~(\ref{eq_4cMS}), from the $c_i/M_N$ group.
This attractive term receives a factor nine in $^1F_3$,
a factor $(-3)$ in $^3G_5$, and a factor one in $^3F_3$
and $^3F_4$. Thus, this contribution is very attractive in
$^1F_3$ and repulsive in $^3G_5$. The latter is the reason for
the overcompensation of the $c_i^2$ graph by
the $c_i/M_N$ contribution in $^3G_5$ which is why the final
N3LO result in this partial wave comes out too repulsive.
One can expect that $1/M_N$ corrections that occur at order
five or six will resolve this problem.

Before finishing this Appendix, we like to point out that
the problem with the $^3G_5$ is not as dramatic as it may appear
from the phase shift plots---for two reasons.
First, the $^3G_5$ phase shifts are about one order of
magnitude smaller than the $F$ and most of the other $G$
phases. Thus, in absolute terms, the discrepancies seen
in $^3G_5$ are small.
In a certain sense, we are looking at `higher order noise'
under a magnifying glass.
Second, 
the $^3G_5$ partial wave contributes 0.06 MeV
to the energy per nucleon in nuclear matter,
the total of which is $-16$ MeV.
Consequently, small discrepancies in the reproduction
of $^3G_5$ by a $NN$ interaction model will have
negligible influence on the microscopic nuclear structure
predictions obtained with that model.

\newpage

\begin{table}
\caption{Parameters used in our calculations. The LECs $c_i$ and $\bar{d}_i$
 are in units of GeV$^{-1}$ and GeV$^{-2}$, respectively.}
\begin{tabular}{cdc}
 & Our choice & Empirical \\
\hline
$M_N$ & 938.9182 MeV  \\
$m_\pi$ & 138.039 MeV   \\
$g_A$ & 1.29 & $1.29\pm 0.01^a$ \\
$f_\pi$ & 92.4 MeV & $92.4\pm 0.3$ MeV$^b$ \\
$c_1$ & --0.81 & $-0.81\pm 0.15^c$ \\
$c_2$ & 3.28 & $3.28\pm 0.23^d$ \\
$c_3$ & --3.40 & $-4.69\pm 1.34^c$ \\
$c_4$ & 3.40 & $3.40\pm 0.04^c$ \\
$\bar{d}_1 + \bar{d}_2$ & 3.06 & $3.06\pm 0.21^d$ \\
$\bar{d}_3$ & --3.27 & $-3.27\pm 0.73^d$ \\
$\bar{d}_5$ & 0.45 & $0.45\pm 0.42^d$ \\
$\bar{d}_{14} - \bar{d}_{15}$ & --5.65 & $-5.65\pm 0.41^d$
\end{tabular}
$^a$Using $g_{\pi NN}^2/4\pi = 13.63\pm 0.20$~\cite{STS93,AWP94} and
applying the Goldberger-Treiman relation,
$g_A  =  g_{\pi NN} \; f_\pi/M_N$.\\ 
$^b$Ref.~\cite{PDG00}.\\
$^c$Table~1, Fit~1 of Ref.~\cite{BM00}.\\
$^d$Table~2, Fit~1 of Ref.~\cite{FMS98}.
\end{table}

\begin{figure}
\vspace{-2cm}
\hspace{1.75cm}
\psfig{figure=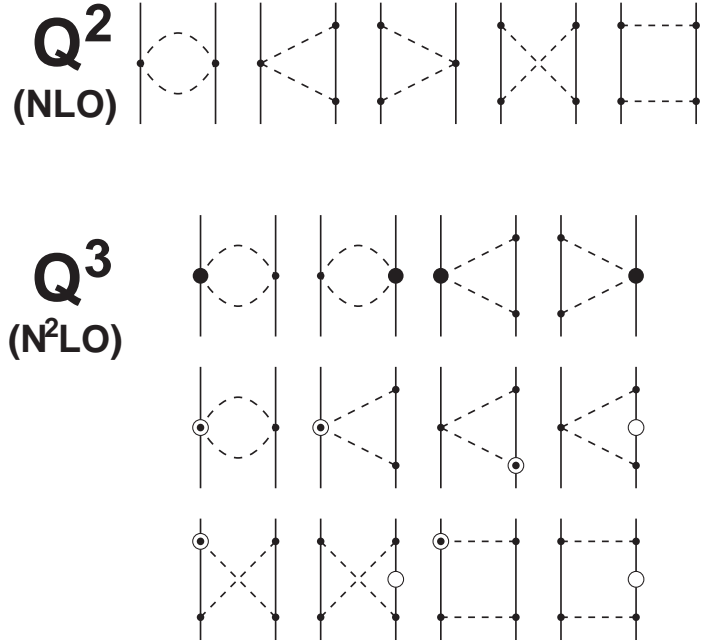,height=17cm}
\vspace{-5cm}
\caption{Two-pion exchange contributions to the $NN$
interaction at order two and three in small momenta.
Solid lines represent nucleons and dashed lines pions.
Small dots denote vertices from the 
leading order $\pi N$ Lagrangian
$\widehat{\cal L}^{(1)}_{\pi N}$,
Eq.~(\protect\ref{eq_L1}). 
Large solid dots are vertices
proportional to the LECs $c_i$
from the second order Lagrangian
$\widehat{\cal L}^{(2)}_{\pi N, \, \rm ct}$,
Eq.~(\protect\ref{eq_L2ct}).
Symbols with an open circles are relativistic $1/M_N$ corrections
contained in the second order Lagrangian
$\widehat{\cal L}^{(2)}_{\pi N}$,
Eqs.~(\protect\ref{eq_L2}).
Only a few representative examples of $1/M_N$ corrections are
shown and not all.
}
\label{fig_diag1}
\end{figure}

\begin{figure}
\vspace{-1cm}
\hspace{4.0cm}
\psfig{figure=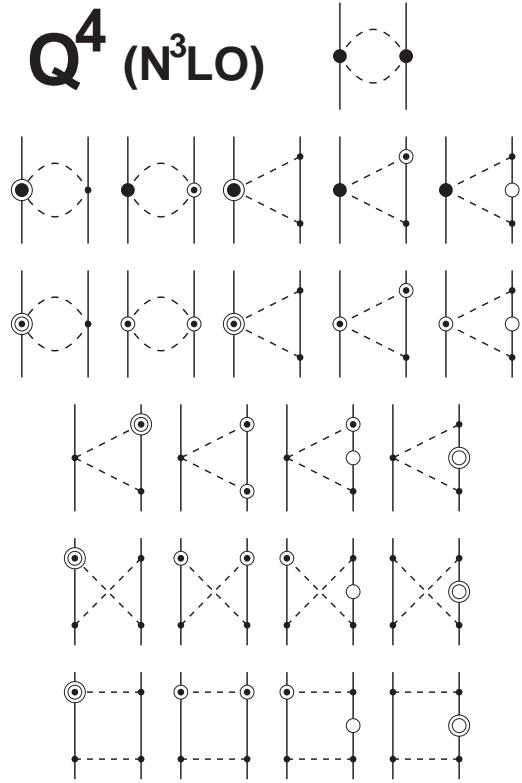,height=15cm}
\vspace{-3.00cm}
\caption{One-loop $2\pi$-exchange contributions to the $NN$
interaction at order four. Basic notation as in Fig.~\ref{fig_diag1}.
Symbols with a large solid dot and an open circle denote 
$1/M_N$ corrections of vertices
proportional to $c_i$.
Symbols with two open circles mark relativistic $1/M^2_N$ corrections.
Both corrections are part of the third order Lagrangian
$\widehat{\cal L}^{(3)}_{\pi N}$, Eq.~(\ref{eq_L3}).
Representative examples for all types of one-loop graphs that occur 
at this order are shown.}
\label{fig_diag2}
\end{figure}

\begin{figure}
\vspace{-1cm}
\hspace{3.5cm}
\psfig{figure=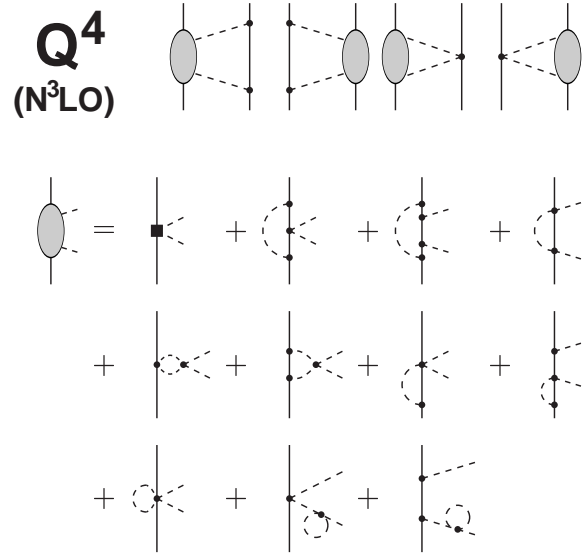,height=15cm}
\vspace{-6.0cm}
\caption{Two-loop $2\pi$-exchange contributions at order four.
Basic notation as in Fig.~\ref{fig_diag1}.
The grey disc stands for all one-loop $\pi N$ graphs
some of which are shown in the lower part of the figure.
The solid square represents vertices proportional to the LECs
$d_i$ which are introduced by the third order Lagrangian
${\cal L}^{(3)}_{\pi N}$,
Eq.~(\ref{eq_L3rel}). More explanations are given in the text.}
\label{fig_diag3}
\end{figure}

\begin{figure}
\vspace*{-3cm}
\hspace*{3.0cm}
\psfig{figure=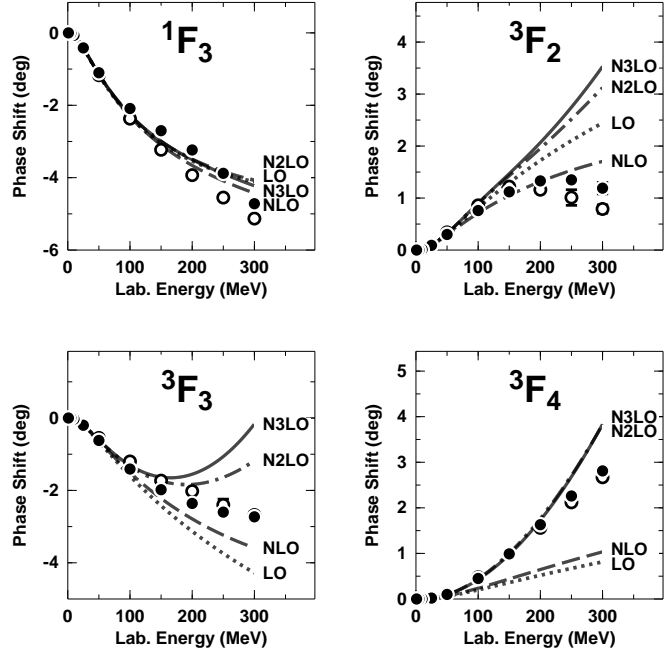,height=14cm}
\vspace*{-2.0cm}
\caption{$F$-wave phase shifts of neutron-proton scattering
for laboratory kinetic energies below 300 MeV.
We show the predictions from chiral pion exchange
to leading order (LO), next-to-leading order (NLO),
next-to-next-to-leading order (N2LO), and
next-to-next-to-next-to-leading order (N3LO).
The solid dots and open circles are the results from the Nijmegen
multi-energy $np$ phase shift analysis~\protect\cite{Sto93} and the VPI
single-energy $np$ analysis SM99~\protect\cite{SM99}, respectively.}
\label{fig_f}
\end{figure}

\begin{figure}
\vspace*{-3cm}
\hspace*{3.0cm}
\psfig{figure=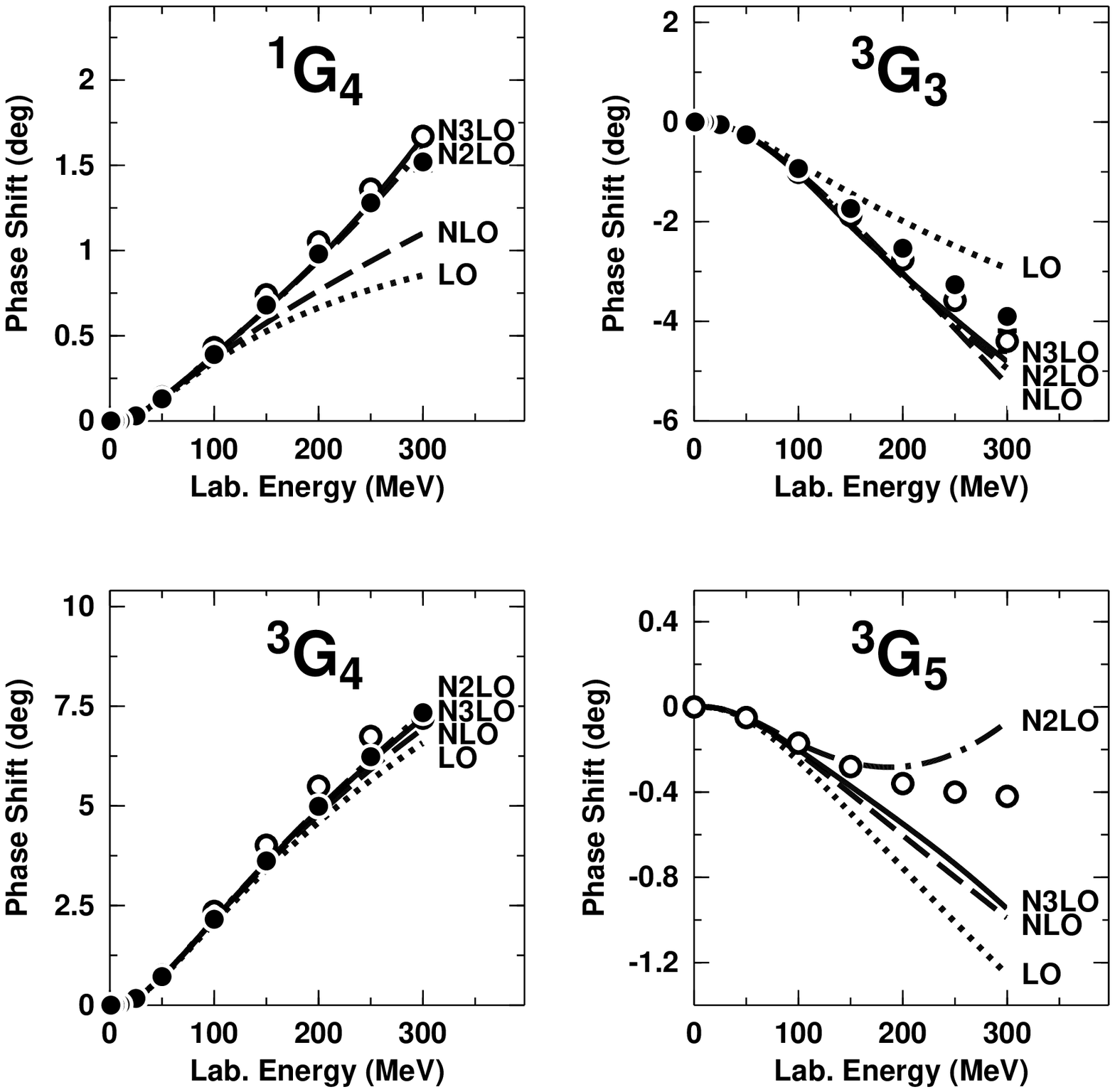,height=14cm}
\vspace*{-2.0cm}
\caption{Same as Fig.~\protect\ref{fig_f},
but for $G$-waves.}
\label{fig_g}
\end{figure}

\begin{figure}
\vspace*{-3cm}
\hspace*{3.0cm}
\psfig{figure=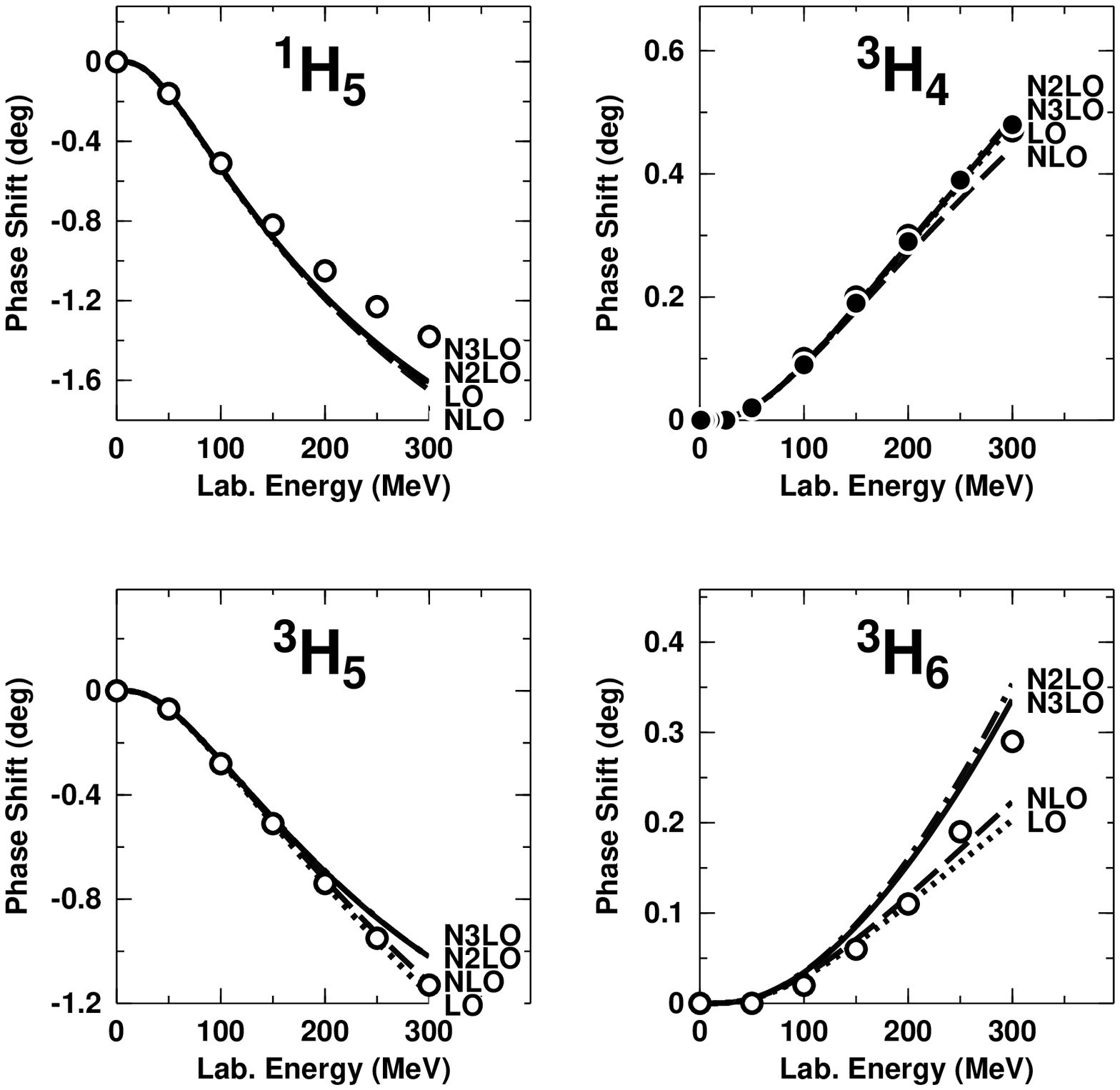,height=14cm}
\vspace*{-2.0cm}
\caption{Same as Fig.~\protect\ref{fig_f},
but for $H$-waves.}
\label{fig_h}
\end{figure}

\begin{figure}
\vspace*{-3cm}
\hspace*{3.0cm}
\psfig{figure=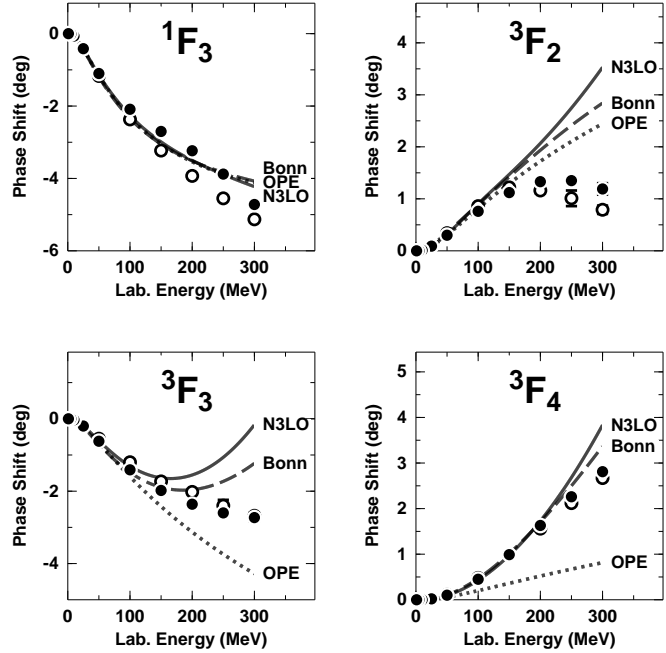,height=14cm}
\vspace*{-2.0cm}
\caption{$F$-wave phase shifts of neutron-proton scattering
for laboratory kinetic energies below 300 MeV.
We show the results from one-pion-exchange (OPE),
and one- plus two-pion exchange as predicted by $\chi$PT at
next-to-next-to-next-to-leading order (N3LO) and 
by the Bonn Full Model~\protect\cite{MHE87} (Bonn).
Empirical phase shifts (solid dots and open circles)
as in Fig.~\protect\ref{fig_f}.}
\label{fig_ff}
\end{figure}

\begin{figure}
\vspace*{-3cm}
\hspace*{3.0cm}
\psfig{figure=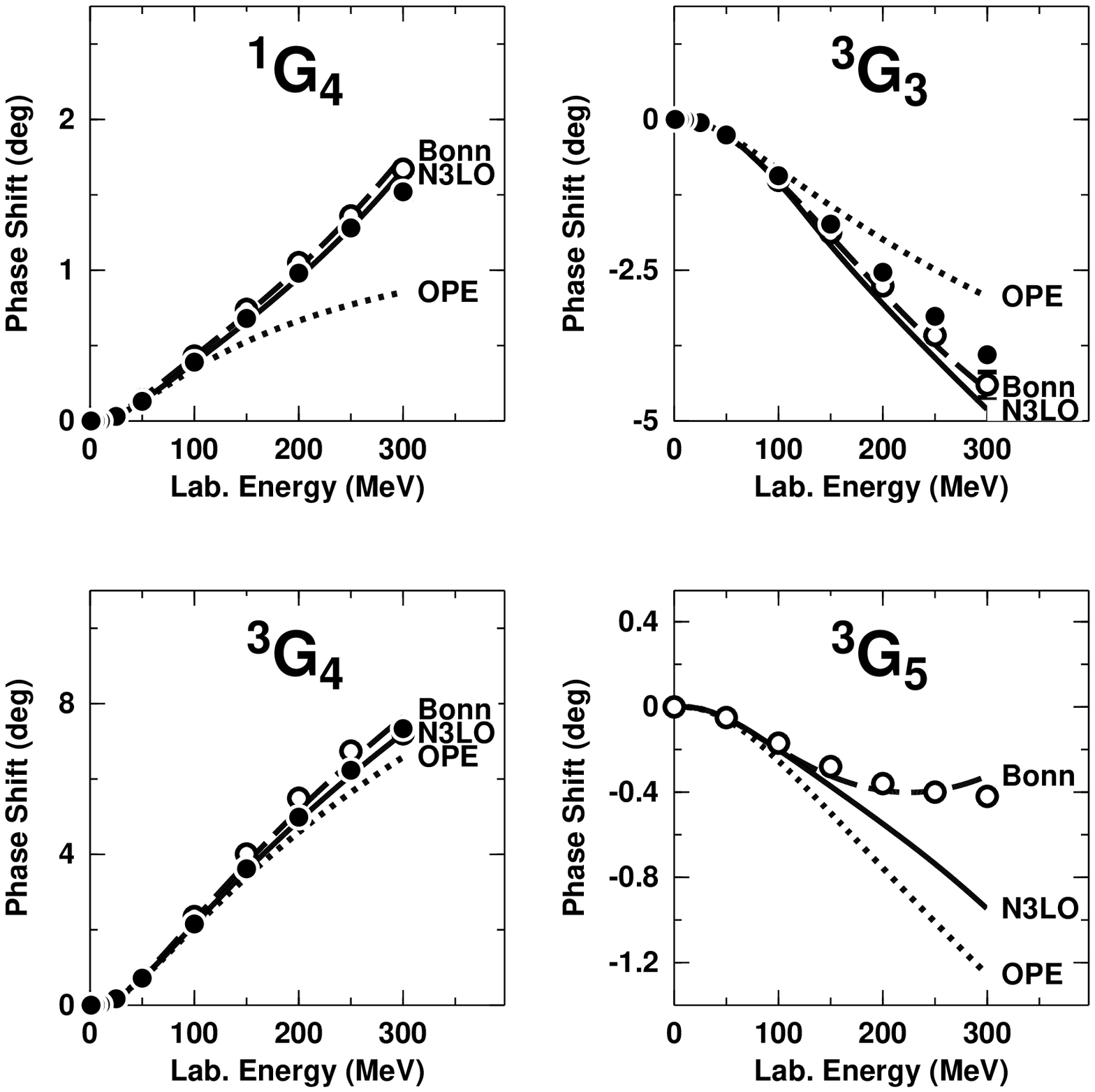,height=14cm}
\vspace*{-2.0cm}
\caption{Same as Fig.~\protect\ref{fig_ff},
but for $G$-waves.}
\label{fig_gg}
\end{figure}

\begin{figure}
\vspace*{-3cm}
\hspace*{3.0cm}
\psfig{figure=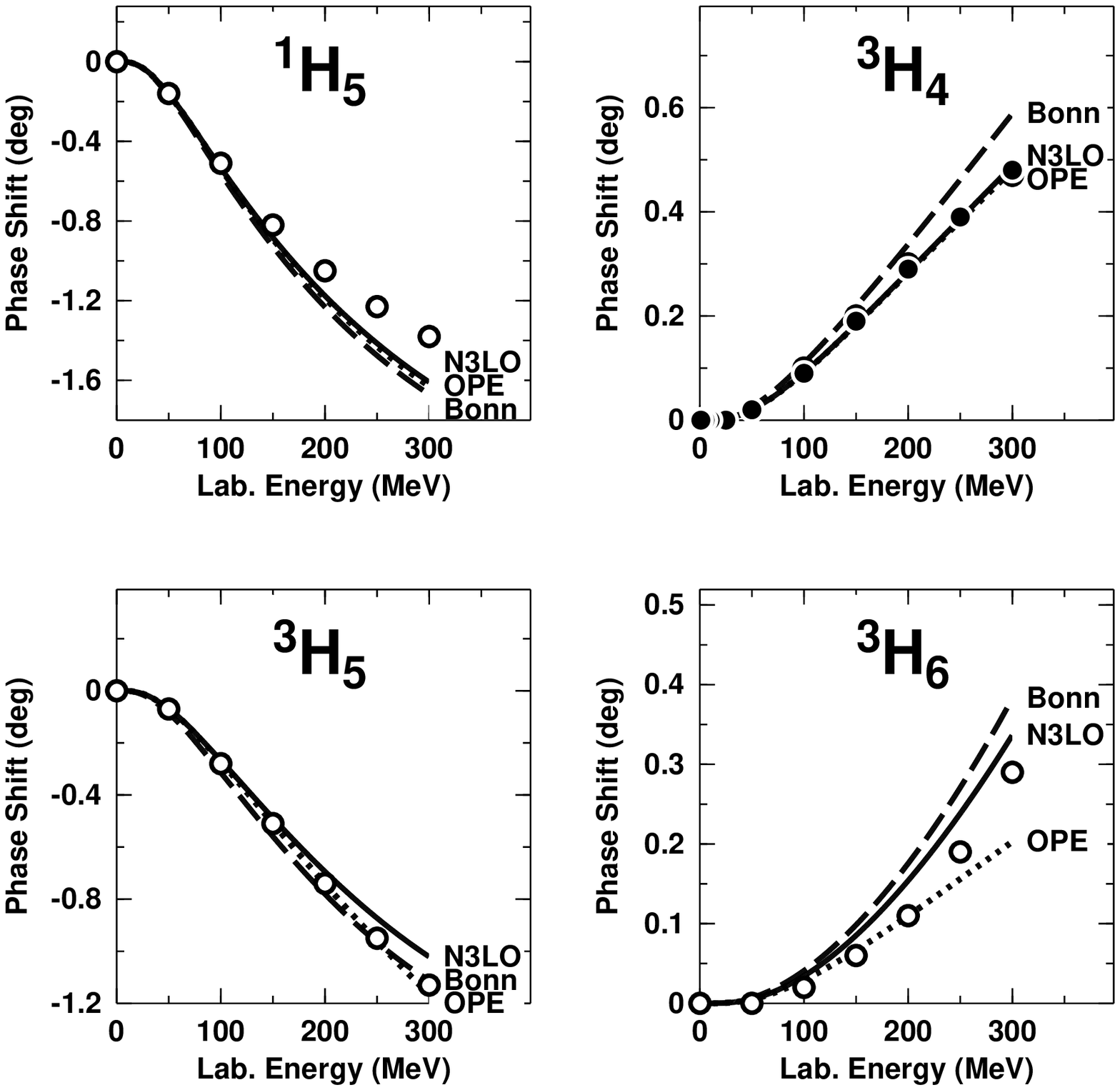,height=14cm}
\vspace*{-2.0cm}
\caption{Same as Fig.~\protect\ref{fig_ff},
but for $H$-waves.}
\label{fig_hh}
\end{figure}

\pagebreak

\begin{figure}
\vspace{-4cm}

\hspace{1cm}\psfig{figure=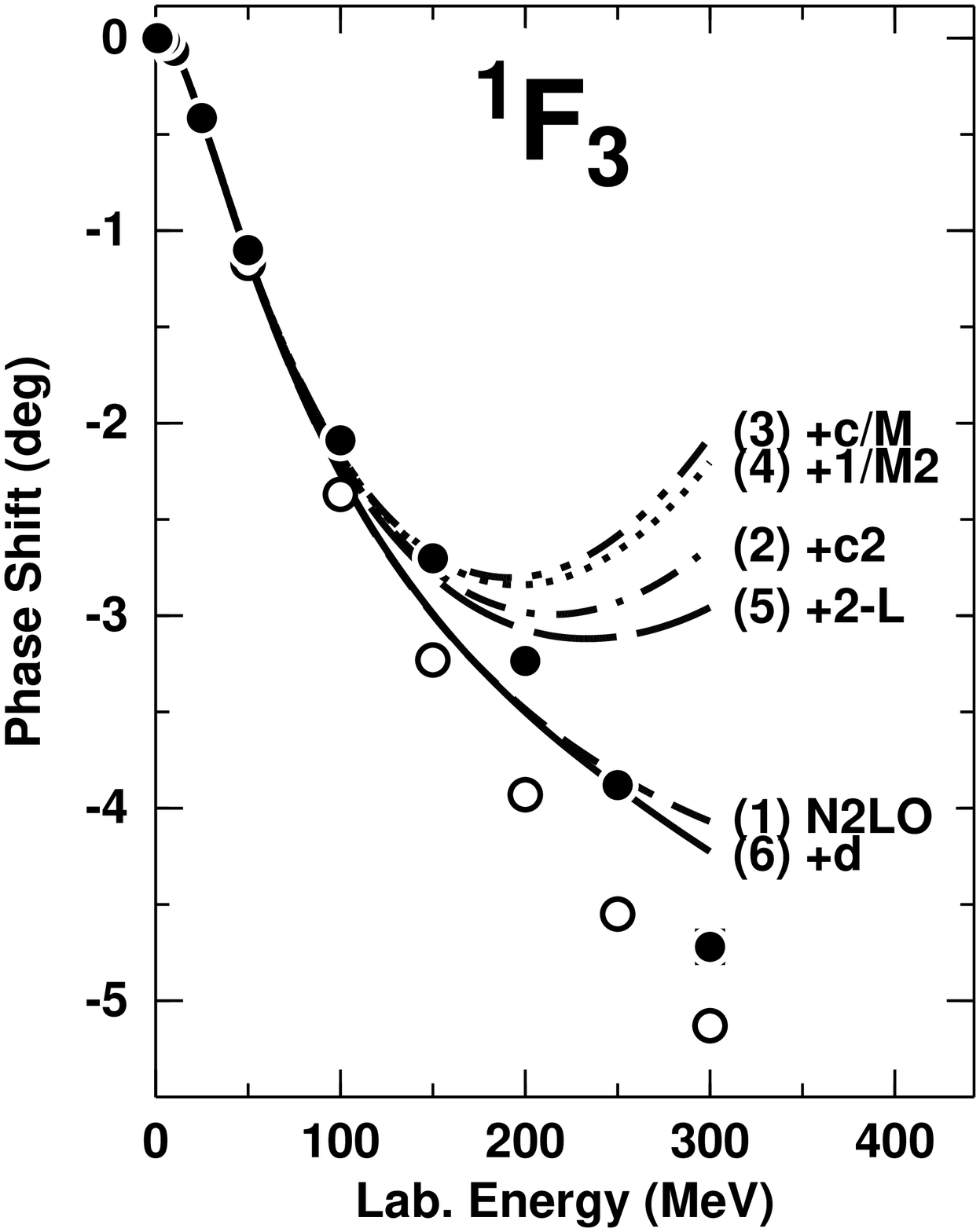,height=10cm}

\vspace{-10cm}\hspace{9cm}\psfig{figure=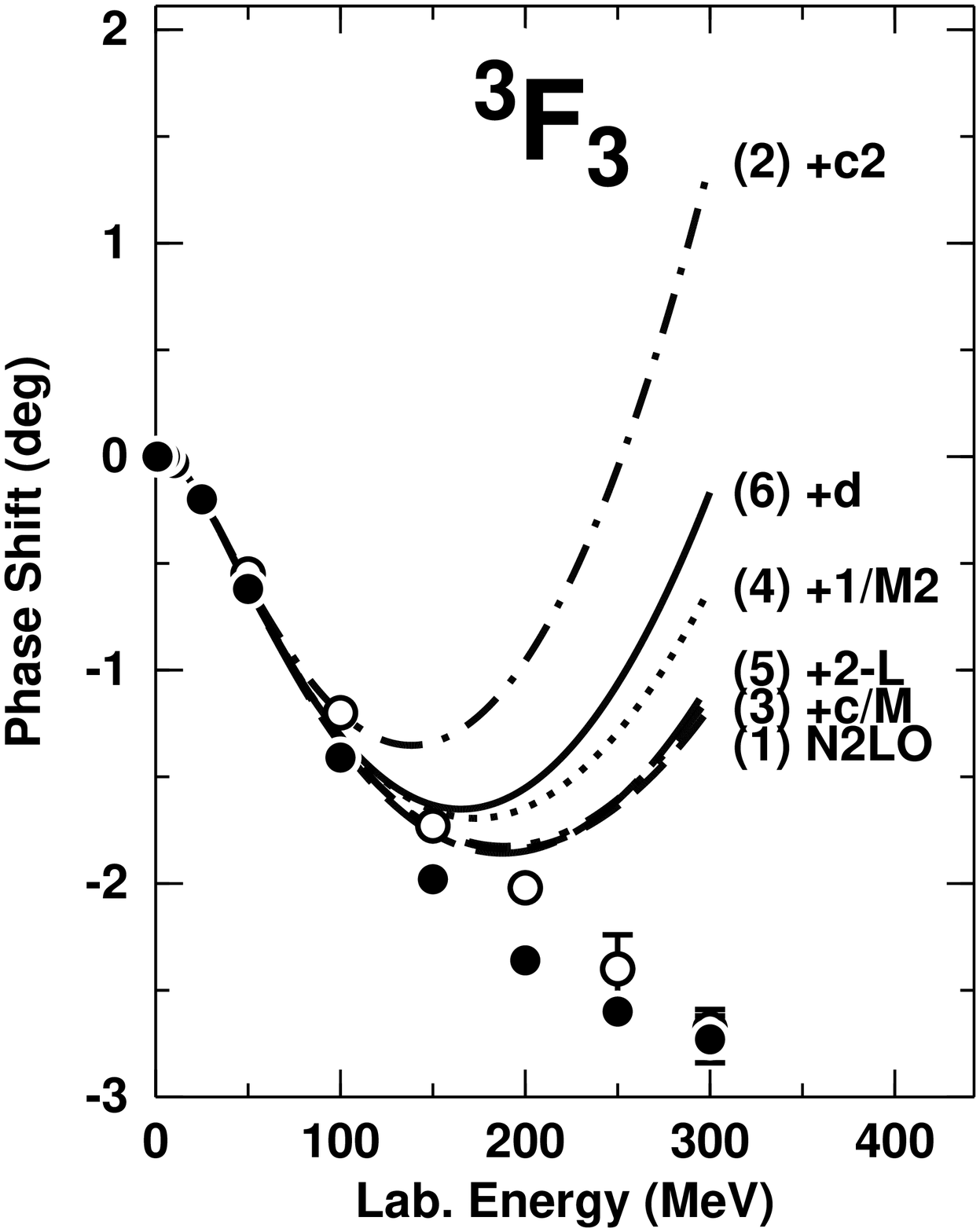,height=10cm}

\vspace{-1cm}

\hspace{1cm}\psfig{figure=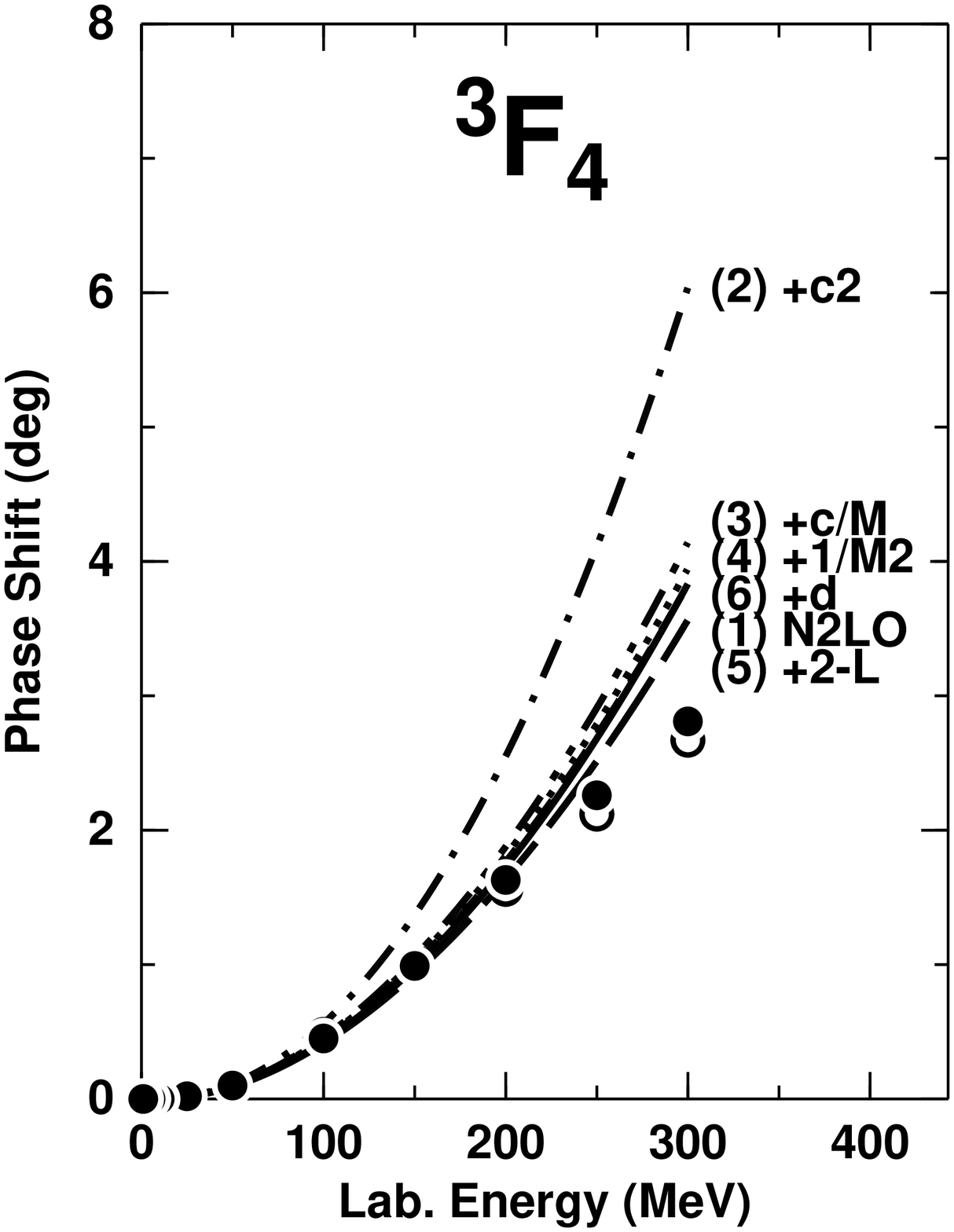,height=10cm}

\vspace{-10cm}\hspace{9cm}\psfig{figure=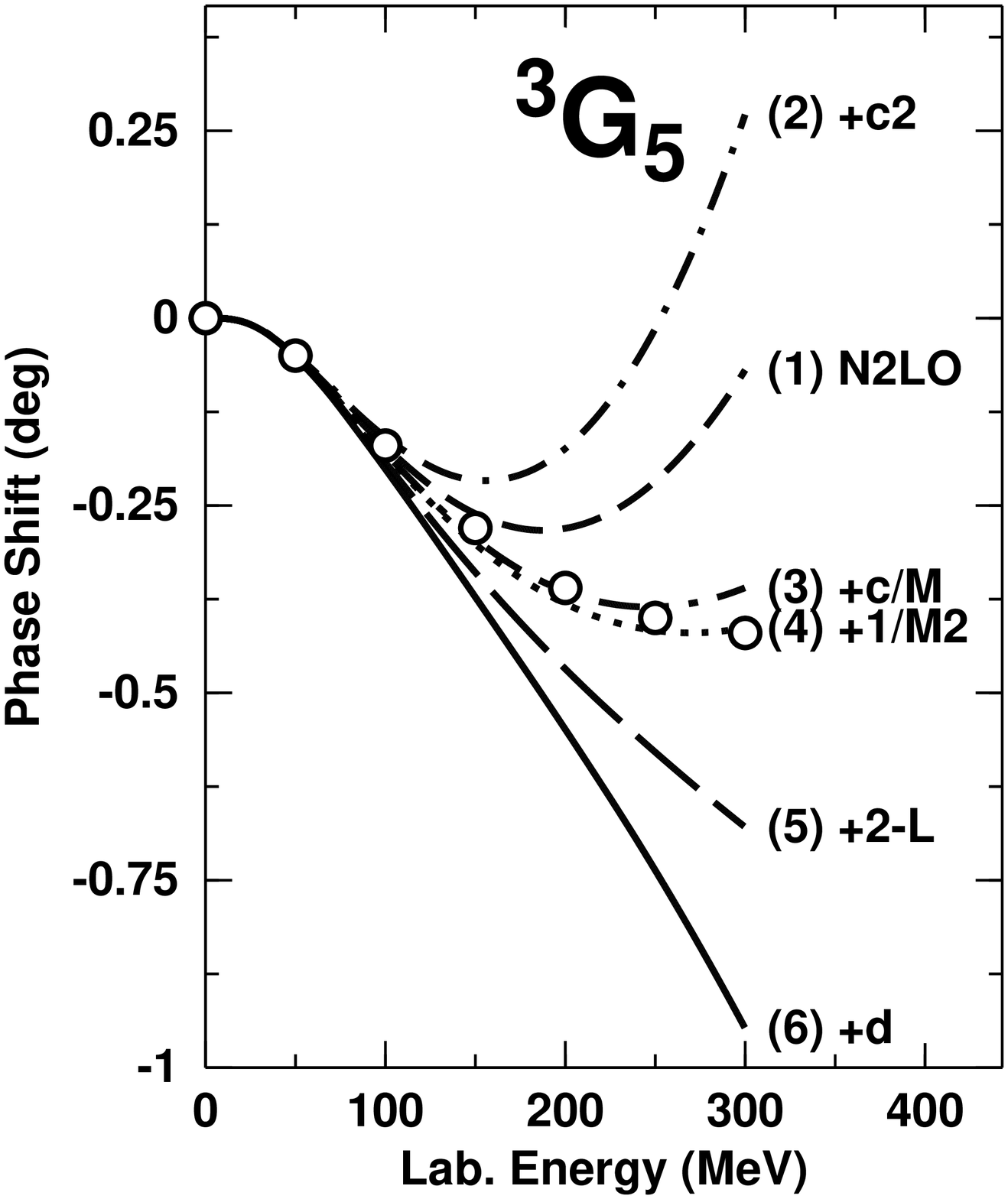,height=10cm}

\vspace{-0.5cm}
\caption{The effect of individual order-four contributions
on the neutron-proton phase shifts in some selected peripheral
partial waves. The individual contributions
are added up successively in the order given in 
parenthesis next to each curve.
Curve (1) is N2LO and curve (6) is the complete N3LO.
For further explanations, see Appendix A.
Empirical phase shifts (solid dots and open circles)
as in Fig.~\protect\ref{fig_f}.}
\label{fig_1f3}
\end{figure}

\begin{figure}
\vspace*{-5cm}
\hspace*{3.0cm}
\psfig{figure=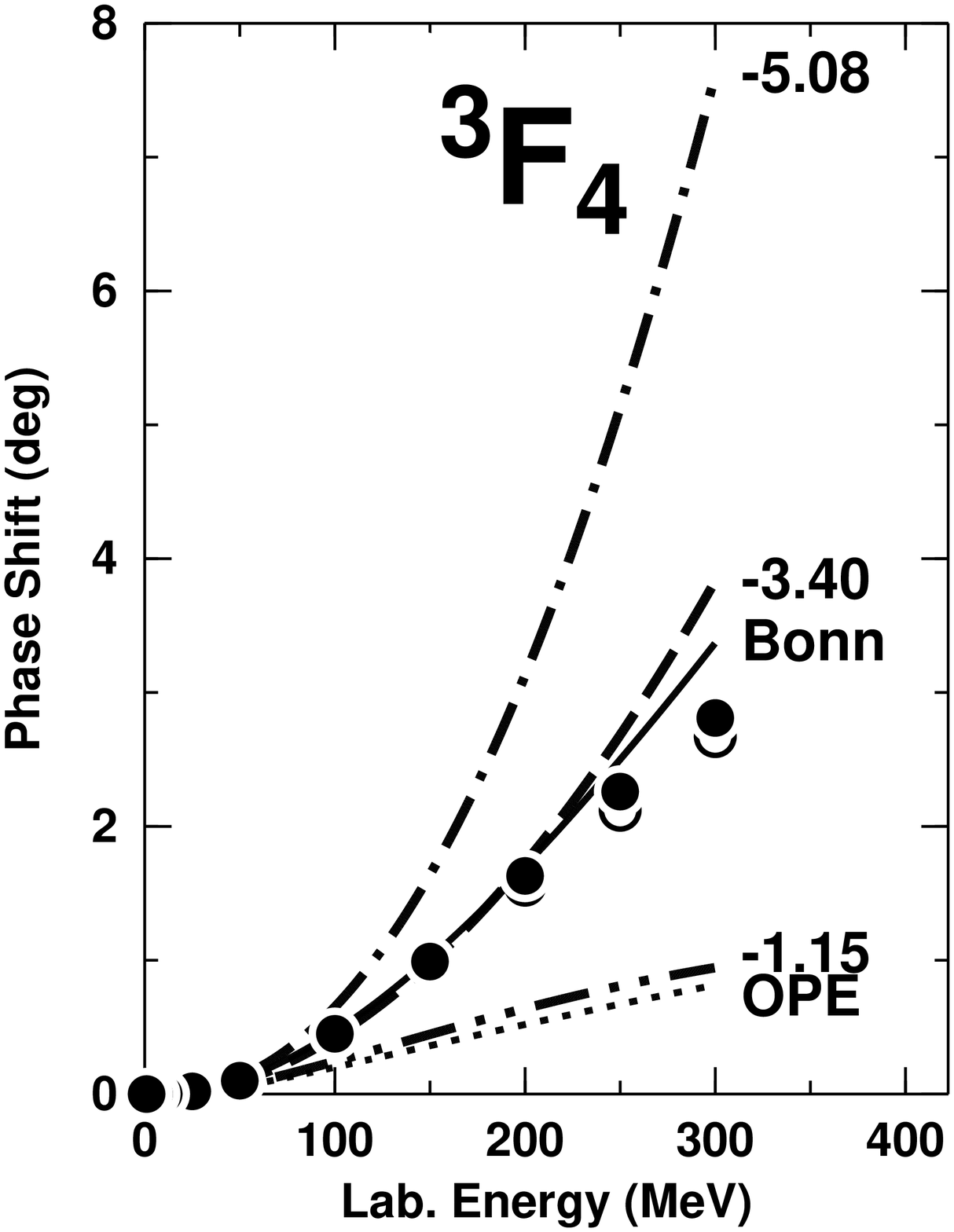,height=14cm}
\vspace*{-0.5cm}
\caption{One and two-pion exchange contributions at order four to the
$^3F_4$ phase shifts for three different choices of the LEC $c_3$.
The numbers given next to the curves denote the values for $c_3$ in units
of GeV$^{-1}$ used for the respective curves (all other parameters
as in Table~I). 
For comparison, we also show the OPE contributrion (OPE)
and the result from $\pi$ + $2\pi$ exchange of the Bonn model (Bonn).
Empirical phase shifts (solid dots and open circles)
as in Fig.~\protect\ref{fig_f}.}
\label{fig_3f4}
\end{figure}


\begin{references}
\bibitem{Wei90} S. Weinberg, Phys.\ Lett.\ B {\bf 251}, 288 (1990);
Nucl.\ Phys.\ {\bf B363}, 3 (1991).
\bibitem{OK92}
C. Ord\'o\~nez and U. van Kolck,
Phys.\ Lett.\ B {\bf 291}, 459 (1992).
\bibitem{ORK94}
C. Ord\'o\~nez, L. Ray, and U. van Kolck,
Phys.\ Rev.\ Lett.\ {\bf 72}, 1982 (1994);
Phys.\ Rev.\ C {\bf 53}, 2086 (1996).
\bibitem{Kol99} U. van Kolck, Prog.\ Part.\ Nucl.\ Phys.\ {\bf 43}, 337 (1999).
\bibitem{CPS92} L. S. Celenza, A. Pantziris, and C. M. Shakin,
Phys. Rev. C {\bf 46}, 2213 (1992).
\bibitem{RR94}
C. A. da Rocha and M. R. Robilotta, 
Phys.\ Rev.\ C {\bf 49}, 1818 (1994);
{\it ibid.} {\bf 52}, 531 (1995);
M. R. Robilotta, Nucl. Phys. {\bf A595}, 171 (1995);
M. R. Robilotta, and C. A. da Rocha, 
Nucl. Phys. {\bf A615}, 391 (1997);
J.-L. Ballot, C. A. da Rocha, and M. R. Robilotta, 
Phys. Rev. C {\bf 57}, 1574 (1998).
\bibitem{KBW97} N. Kaiser, R. Brockmann, and W. Weise,
Nucl.\ Phys.\ {\bf A625}, 758 (1997).
\bibitem{KGW98} N. Kaiser, S. Gerstend\"orfer, and W. Weise,
Nucl.\ Phys.\ {\bf A637}, 395 (1998).
\bibitem{Kai99} N. Kaiser, Phys.\ Rev.\ C
{\bf 61}, 014003 (1999);
{\it ibid.\/} {\bf 62}, 024001 (2000).
\bibitem{Kai01} N. Kaiser, Phys.\ Rev.\ C
{\bf 63}, 044010 (2001).
\bibitem{Kai01a} N. Kaiser, Phys.\ Rev.\ C
{\bf 64}, 057001 (2001).
\bibitem{Kai01b}
N. Kaiser, Phys. Rev. C {\bf 65}, 017001 (2002).
\bibitem{EGM98} E. Epelbaum, W. Gl\"ockle, and U.-G. Mei\ss ner,
Nucl.\ Phys.\ {\bf A637}, 107 (1998); {\it ibid.\/}
{\bf A671}, 295 (2000).
\bibitem{EM01} D. R. Entem and R. Machleidt,
Phys. Lett. {\bf B524}, 93 (2002).
\bibitem{KSW96} 
D. B. Kaplan, M.J. Savage, and M.B. Wise, 
Nucl. Phys. {\bf B478}, 629 (1996);
D. B. Kaplan, Nucl. Phys. {\bf B494}, 471 (1997);
D. B. Kaplan, M.J. Savage, and M.B. Wise, 
Nucl. Phys. {\bf B534}, 329 (1998);
Phys. Lett. {\bf B424}, 390 (1998).
\bibitem{FST97} R. J. Furnstahl, B. D. Serot, and H.-B. Tang,
Nucl. Phys. {\bf A615}, 441 (1997);
J. V. Steel and R. J. Furnstahl, 
{\it ibid.} {\bf A637}, 46 (1998);
R. J. Furnstahl, J. V. Steel, and N. Tirfessa, 
{\it ibid.} {\bf A671}, 396 (2000);
R. J. Furnstahl, H. W. Hammer, and N. Tirfessa, 
{\it ibid.} {\bf A689}, 846 (2001).
\bibitem{Par98} 
T.-S. Park, K. Kubodera, D. P. Min, and M. Rho,
Phys. Rev. C {\bf 58}, 637 (1998);
Nucl. Phys. {\bf A646}, 83 (1999).
\bibitem{Coh97}
T. D. Cohen, Phys. Rev. C {\bf 55}, 67 (1997);
D. R. Phillips and T. D. Cohen, Phys. Lett. {\bf B390}, 7 (1997);
K. A. Scaldeferri, D. R. Phillips, C. W. Kao, and T. D. Cohen, 
Phys. Rev. C {\bf 56}, 679 (1997);
S. R. Beane, T. D. Cohen, and D. R. Phillips, 
Nucl. Phys. {\bf A632}, 445 (1998).
\bibitem{RS99}
G. Rupak and N. Shoresh, nucl-th/9906077;
P. F. Bedaque, H. W. Hammer, and U. van Kolck,
Nucl. Phys. {\bf A676}, 357 (2000);
S. Fleming, Th. Mehen, and I. W. Stewart,
Nucl. Phys. {\bf A677}, 313 (2000);
Phys. Rev. C {\bf 61}, 044005 (2000).
\bibitem{Bea01}
S. R. Bean, P. F. Bedaque, M. J. Savage, and U. van Kolck, 
nucl-th/0104030.
\bibitem{Ren99} M. C. M. Rentmeester, R. G. E. Timmermans, J. L. Friar,
and J. J. de Swart, Phys. Rev. Lett. {\bf 82}, 4992 (1999).
\bibitem{FMS98} N. Fettes, U.-G. Mei\ss ner, S. Steiniger,
Nucl.\ Phys.\ {\bf A640}, 199 (1998).
\bibitem{Fet00} N. Fettes, U.-G. Mei\ss ner, M. Moj\v{z}i\v{s}, and S. Steininger,
Ann.\ Phys.\ (N.Y.)\ {\bf 283}, 273 (2000);
{\it ibid.\/} {\bf 288}, 249 (2001).
\bibitem{STS93} V. Stoks, R. Timmermans, and J. J. de Swart,
Phys. Rev. C {\bf 47}, 512 (1993).
\bibitem{AWP94} R. A. Arndt, R. L. Workman, and M. M. Pavan,
Phys. Rev. C {\bf 49}, 2729 (1994).
\bibitem{PDG00} Review of Particle Physics, 
Eur. Phys. J. C {\bf 15}, 1 (2000).
\bibitem{BM00} P. B\"{u}ttiker and U.-G. Mei\ss ner,
Nucl.\ Phys.\ {\bf A668}, 97 (2000).
\bibitem{BKM95} V. Bernard, N. Kaiser, and U.-G. Mei\ss ner,
Int.\ J.\ Mod.\ Phys.\ E {\bf 4}, 193 (1995).
\bibitem{LM98} G. Q. Li and R. Machleidt, Phys. Rev. C {\bf 58},
3153 (1998).
\bibitem{JW59} M. Jacob and G. C. Wick, Ann. Phys. (N.Y.) {\bf 7}, 404 (1959).
\bibitem{GM61} J. Goto and S. Machida, Prog.\ Theor.\ Phys.\
{\bf 25}, 64 (1961).
\bibitem{EAH71} K. Erkelenz, R. Alzetta, and K. Holinde, Nucl. Phys. {\bf 
A176}, 413 (1971);
note that there is
an error in equation~(4.22) of this paper where it should read:
$^-W^J_{LS}=2qq'\frac{J-1}{2J-1} [A^{J-2,(0)}_{LS} - A^{J(0)}_{LS}]$
and
$^+W^J_{LS}=2qq'\frac{J+2}{2J+3} [A^{J+2,(0)}_{LS} - A^{J(0)}_{LS}]$.
\bibitem{SYM57} H. P. Stapp, T. J. Ypsilantis, and N. Metropolis, Phys. Rev. 
{\bf 105}, 302 (1957).
\bibitem{Sto93} V.\ G.\ J.\ Stoks, R.\ A.\ M.\ Klomp, M.\ C.\ M.\ Rentmeester,
and J.\ J.\ de Swart, Phys.\ Rev.\ C {\bf 48}, 792 (1993).
\bibitem{SM99} 
R. A. Arndt, I. I. Strakovsky, and R. L. Workman,
SAID, Scattering Analysis Interactive Dial-in computer facility,
Virginia Polytechnic Institute and George Washington University,
solution SM99 (Summer 1999); for more information see, e.~g.,
R. A. Arndt, I. I. Strakovsky, and R. L. Workman,
Phys. Rev. C {\bf 50}, 2731 (1994).
\bibitem{MHE87} R. Machleidt, K. Holinde, and Ch. Elster,
Phys.\ Rep.\ {\bf 149}, 1 (1987).
\bibitem{foot3} Note that the Bonn model uses the $\pi NN$ coupling
constant 
$g_{\pi NN}^2/4\pi = 14.4$, while for chiral pion exchanges
we apply
$g_{\pi NN}^2/4\pi = 13.6$ [cf.\ footnote $^a)$ of Table~I].
This is the main reason for the light ``discrepancies'' between
N3LO and Bonn that seem to show up in some of the $H$ waves
(Fig.~\ref{fig_hh}).
\bibitem{foot2}
In fact, preliminary calculations, which take an important
class of diagrams of order five into account, indicate that
the N$^4$LO contribution may prevailingly be repulsive
(N. Kaiser, private communication).
\bibitem{BJ76} 
G. E. Brown and A. D. Jackson, {\it The Nucleon-Nucleon
Interaction}, (North-Holland, Amsterdam, 1976).
\bibitem{Paris} 
R. Vinh Mau, in: {\it Mesons in Nuclei}, Vol.~I, 
M. Rho and D. H. Wilkinson, eds. 
(North-Holland, Amsterdam, 1979)
p.~151;
M. Lacombe, B. Loiseau, J. M. Richard, R. Vinh Mau,
J. C\^{o}t\'{e}, P. Pires, and R. de Tourreil, 
Phys. Rev. C {\bf 21}, 861 (1980).
\bibitem{Epe02}
E. Epelbaum, A. Nogga, W. Gl\"ockle, H. Kamada, U.-G. Mei\ss ner,
and H. Witala, {\it Few-Nucleon Systems with Two-Nucleon Forces
from Chiral Effective Field Theory}, {\tt nucl-th/0201064}.
\bibitem{CDR72} M. Chemtob, J. W. Durso, and D. O. Riska,
Nucl. Phys. {\bf B38}, 141 (1972);
A. D. Jackson, D. O. Riska, and B. Verwest,
Nucl. Phys. {\bf A249}, 397 (1975).
\bibitem{Vin73} R. Vinh Mau, J. M. Richard, B. Loiseau,
M. Lacombe, and W. M. Cottingham, Phys. Lett. {\bf B44}, 1 (1973);
M. Lacombe, B. Loiseau, J. M. Richard, R. Vinh Mau,
P. Pires, and R. de Tourreil, 
Phys. Rev. D {\bf 12}, 1495 (1975).
\bibitem{foot} We note that in the Nijmegen analysis~\cite{Ren99}
$c_4=4.7\pm 0.7$ Gev$^{-1}$ emerges which also differs from what
we use and what is obtained in $\pi N$ analysis (cf.\ Table~I).
However, this difference in $c_4$ has very little 
impact on $NN$ peripheral partial waves and does not make possible
stronger values for $c_3$.
\end{references}
\end{document}